\documentclass[submission, Phys]{SciPost}

\usepackage{orcidlink}
\usepackage{amssymb}
\begin{document}

\begin{center}{\Large \textbf{
Magnetic field – bias current interplay in HgTe-based three-terminal Josephson junctions
}}\end{center}

\begin{center}
J. Thieme\textsuperscript{\orcidlink{0009-0008-5538-6891}1*$\dagger$},
W. Himmler  \textsuperscript{\orcidlink{0000-0002-8706-4626}1*},
F. Dominguez \textsuperscript{\orcidlink{0000-0003-1400-2165}2*},
G. Platero \textsuperscript{\orcidlink{0000-0001-8610-0675}3},
N. Hüttner \textsuperscript{\orcidlink{0000-0002-7703-313X}1},
S. Hartl \textsuperscript{\orcidlink{0009-0003-3717-1487}1}
E. Richter \textsuperscript{\orcidlink{0009-0006-9611-3350}1}
D. A. Kozlov \textsuperscript{\orcidlink{0000-0003-0639-8980}1}
N. N. Mikhailov \textsuperscript{\orcidlink{0009-0007-7474-102X}4}
S. A. Dvoretsky \textsuperscript{\orcidlink{0000-0002-1295-5598}4}
D. Weiss \textsuperscript{\orcidlink{0000-0002-9630-9787}1}
\end{center}

\begin{center}
{\bf 1} Institute of Experimental and Applied Physics, University of Regensburg, D-93040 Regensburg, Germany
\\
{\bf 2} Institute for Theoretical Physics and Astrophysics, and W\"urzburg-Dresden Cluster of Excellence on Complexity, Topology and Dynamics in Quantum Matter ctd.qmat, Julius-Maximilians-Universit\"at W\"urzburg, Am Hubland, D-97074 W\"urzburg, Germany
\\
{\bf 3} Instituto de Ciencia de Materiales de Madrid ICMM-CSIC, 28049 Madrid, Spain
\\
{\bf 4} A. V. Rzhanov Institute of Semiconductor Physics, Novosibirsk 630090, Russia
\\
* These authors contributed equally to this work. 
\\
$\dagger$ Corresponding author j.thieme@fz-juelich.de
\end{center}

\begin{center}
\today
\end{center}

\section*{Abstract}
{\bf
We investigate HgTe/Nb-based three-terminal Josephson junctions in T-shaped and X-shaped geometries and their critical current contours (CCCs). By decomposing the CCCs into the contributions from individual junctions, we uncover how bias current and magnetic field jointly determine the collective Josephson behavior. A perpendicular magnetic field induces a tunable crossover between SQUID-like and Fraunhofer-like interference patterns, controlled by the applied bias. Moreover, magnetic flux produces pronounced deformations of the CCC, enabling symmetry control in the 
$(I_1,I_2)$ plane. Remarkably, we identify a regime of strongly enhanced Josephson diode efficiency, reaching values up to $\eta\approx 0.8$ at low bias and magnetic field. The experimental results are quantitatively reproduced by resistively shunted junction (RSJ) simulations, which capture the coupled dynamics of current and flux in these multi-terminal superconducting systems.
}

\vspace{10pt}
\noindent\rule{\textwidth}{1pt}
\tableofcontents\thispagestyle{fancy}
\noindent\rule{\textwidth}{1pt}
\vspace{10pt}

\section{Introduction}
\label{sec:intro}

Multiterminal Josephson junctions (MTJJs) provide a powerful platform for realizing and controlling topological states of matter~\cite{Heck2014a, Yokoyama2015a, riwar_multi-terminal_2016, Meyer2017a, Xie2017a, Eriksson2017a}. In these systems, the superconducting phase differences between multiple terminals play a role analogous to that of crystal momentum in a Brillouin zone, allowing the definition of a synthetic Chern number in phase-difference space~\cite{Meyer2017a, Xie2018a, meyer_conductance_2021}. This framework enables Andreev bound states to host a variety of topological phases, such as Weyl singularities in four-terminal junctions~\cite{riwar_multi-terminal_2016, Repin2019a} and Majorana bound states~\cite{Gavensky2019a, Houzet2019a, Trif2019a, Sakurai2020a}. Furthermore, MTJJs even allow for the measurement of the quantum geometric tensor~\cite{Klees2020a, Klees2021a}, establishing them as a versatile setting for realizing and probing topological quantum states.

Beyond their connection to topological states of matter, multiterminal Josephson junctions also hold promise for novel superconducting circuit applications. 
They can exhibit an inherent superconducting diode effect, that is, a non-resistive current when the junction is biased in one direction and a resistive response when the bias is applied in the opposite direction~\cite{gupta_gate-tunable_2023, chiles_nonreciprocal_2023, behner_superconductive_2025, correa_theory_2024}. 
One of the main advantages of using MTJJs, compared to conventional Josephson junctions, arises from the inherent breaking of time-reversal and inversion symmetries, which is required to observe this effect. Here, time-reversal symmetry is broken by the applied bias currents, while inversion symmetry is broken by the geometry of the junction.

In the past decade, the first MTJJs have been realized in T-, Y- and X-shaped geometries (with two contacts shorted), primarily based on semi-metals and a few topological insulators (TIs)~\cite{graziano_transport_2020,kolzer_-plane_2021,kolzer_supercurrent_2023,gupta_gate-tunable_2023,pankratova_multiterminal_2020,graziano_selective_2022, chiles_nonreciprocal_2023, arnault_dynamical_2022,zhang_anomalous_2022, behner_superconductive_2025}.
These initial characterizations of three-and four-terminal devices have provided valuable insight into their fascinating properties. They also demonstrate the wide parameter space that becomes accessible by tuning the gauge-invariant phase difference between different superconducting leads. So far, research has mainly focused on the effects of magnetic fields \cite{pankratova_multiterminal_2020, gupta_gate-tunable_2023}, gate dependence \cite{graziano_transport_2020,arnault_dynamical_2022,graziano_selective_2022,chiles_nonreciprocal_2023}, and diode effects \cite{gupta_gate-tunable_2023,chiles_nonreciprocal_2023,behner_superconductive_2025} in fully coupled junctions. However, while the entire system has been investigated, the relation to its individual-junction components remains largely unexplored.

In this work, we investigate three-terminal Josephson junctions based on HgTe and show that the critical current contour (CCC) is highly tunable and sensitively depend on the combined action of magnetic field and bias currents. By systematically probing their behavior as a function of these parameters, we identify a crossover between distinct interference regimes, ranging from SQUID-like to Fraunhofer-like patterns, whose characteristics can be continuously controlled by the applied current configuration. This interplay results in pronounced deformations and symmetry breaking of the CCC in the multidimensional current space, reflecting the nontrivial redistribution of supercurrents among the terminals. As a direct consequence of the combined breaking of spatial and time-reversal symmetries, we observe a large and tunable Josephson diode effect with high efficiency. These findings are quantitatively captured by a resistively shunted junction (RSJ) model incorporating magnetic-field-induced phase modulation, which allows us to disentangle the contributions of the individual junctions and reproduce the observed contour behavior. Our results demonstrate that multiterminal Josephson junctions constitute a versatile platform in which magnetic flux and bias currents act as interdependent control parameters to engineer the supercurrent landscape.


\section{Device parameters and experimental setup}
\label{sec:device}
The devices investigated in this work are based on the topological insulator \text{HgTe}, grown lattice-matched between layers of \text{Cd}$_{0.7}$Hg$_{0.3}$\text{Te} and \text{CdTe} on a \text{GaAs} substrate. For device fabrication, a $15\,\mu \text{m} \times 15\,\mu \text{m}$ mesa structure is etched using a \text{Br}-based wet etching solution \cite{illing_fabrication_1994,ziegler_probing_2018}. To define the MTJJ structure, the capping layers are locally removed with a lower \text{Br} concentration solution. The sample is then immediately transferred into a UHV chamber, where it is gently ion-milled with an \text{Ar}$^+$ plasma to remove surface oxides formed during transfer and to improve the Josephson contact quality with \text{HgTe}. Next, 5\,nm of \text{Ti} are deposited by e-beam evaporation, to serve as an adhesive for the \text{Nb}, which is subsequently sputter-deposited to a thickness of 80\,nm. The leads are capped with a 5\,nm \text{Au} layer (e-beam), to protect the superconductor from oxidation and ensure ohmic contacts. The \text{Nb} contacts exhibit high transparency, with values up to $D \approx 0.7$ and couple to the topological surface states, which form a two-dimensional electron system \cite{Himmler2023_supercurrent_interference, Ziegler2020}. While bulk states may also be present, they are expected to be more strongly suppressed than the surface states and therefore contribute only weakly to the supercurrent \cite{Hartl2025_HgTe_current_distribution}. However, despite the topological character of HgTe, we do not expect any contribution from topological superconductivity in our results. In particular, the dimensions of our devices preclude self-interference effects of the topological surface states, as reported in ~\cite{ziegler_probing_2018, Fischer2022_4pi, Himmler2023_supercurrent_interference}. Furthermore, we do not expect edge transport to play a significant role. Since the area of the topological insulator is much larger than that of the multiterminal junction, proximity-induced superconductivity is confined to the junction region and does not extend throughout the entire sample.

Finally, a stack of 30\,nm \text{SiO}$_2$ (PECVD) and 80\,nm \text{AlOx} (ALD) was deposited globally over the sample, serving both as an oxidation barrier and insulating layer for the \text{Ti}/\text{Au} top gates evaporated to cover the central region of the mesa. During the measurements presented here, however, the top gates were left floating to prevent dielectric break down and the resulting leakage currents.\\
Both samples were measured in an \textit{Oxford Instruments Kelvinox TLM} dilution refrigerator at a base temperature of $\sim20\,\text{mK}$. The cryostat was equipped with a superconducting magnet and a rotatable sample holder, allowing the application of magnetic fields at different angles. DC signals were applied and recorded using a \textit{Nanonis Tramea} system. To reduce measurement time, only DC sweeps were recorded. Whenever possible, a quasi-four-point measurement setup was used, in which two terminals supplied current to the sample through a series resistor, and the third terminal acts as a drain (Fig. \ref{fig1}(a)). Each superconducting terminal was additionally connected to a separate voltage line, allowing measurement of the voltage drop between each pair of contacts. This enabled calculation of the differential resistance of each junction individually, using the measured voltage drop across the single junction and the applied current. 

\section{Results}
\label{sec:exp}

\begin{figure}[h]
\centering
    \includegraphics[width=0.7\textwidth, keepaspectratio] {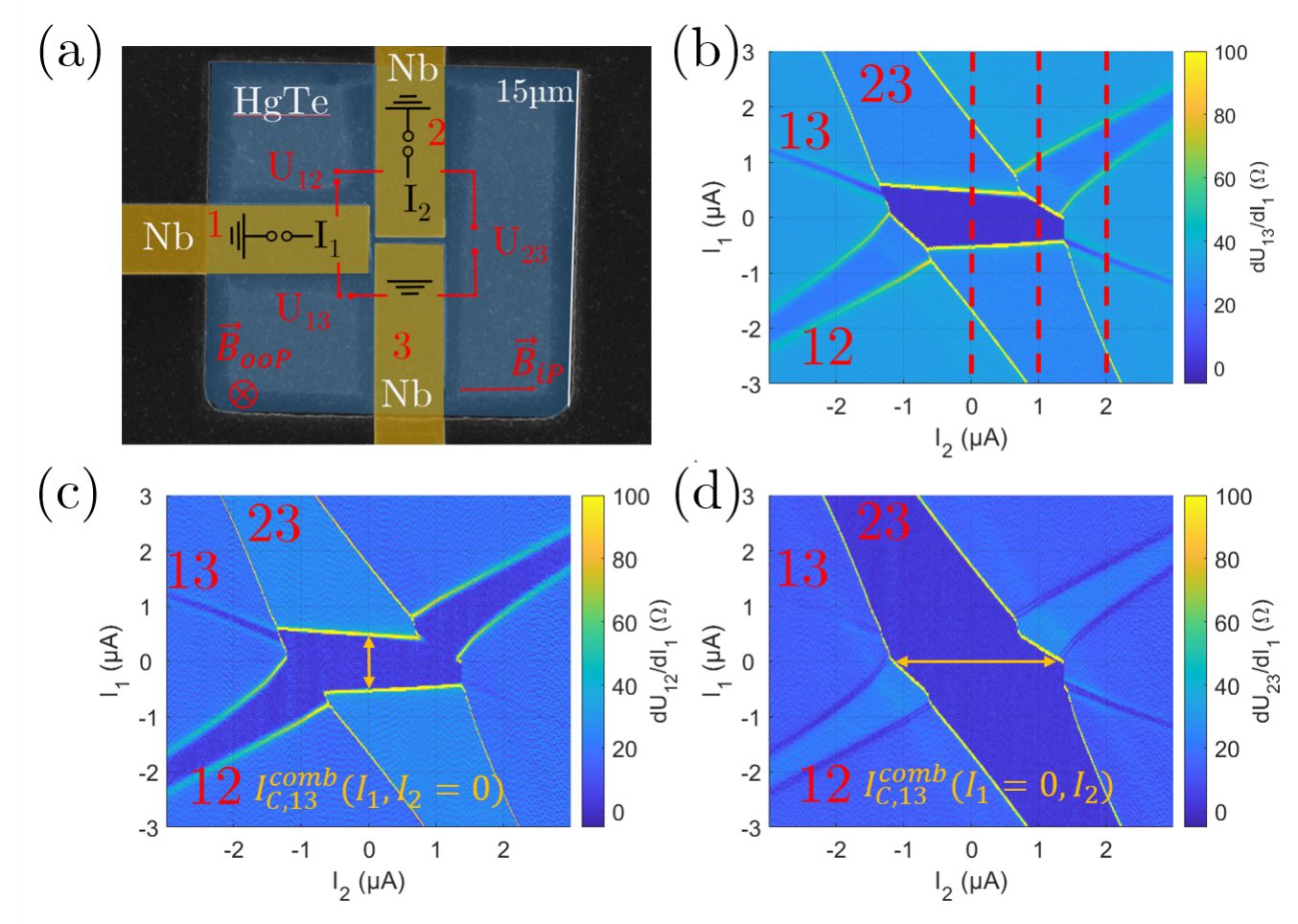}
    \caption{
    \textbf{Overview of the T-device}. 
    (a): Schematic of the measured T-device. The superconducting \text{Nb} shown in yellow and \text{HgTe} in blue. Red numbers indicate the terminals and junctions, along with the corresponding voltages. The current sources are shown in black. Panels (b)-(d) show the differential resistances $dU_{13}/dI_{1}$, $dU_{12}/dI_{1}$ and $dU_{23}/dI_{1}$ of the corresponding junctions JJ$_{13}$, JJ$_{12}$ and JJ$_{23}$ as a function of the bias currents $I_1$ and $I_2$. The currents $I^{\text{comb}}_\text{c,13}(I_1,I_2=0)$ and $I^{\text{comb}}_\text{c,13}(I_1=0,I_2)$ (see section \ref{lbr} in the main text) are highlighted in orange. The vertical red dashed lines in panel (b) indicate line cuts at $I_2=0,\, 1\,\mu \text{A},$ and $2\,\mu$A studied below as a function of an external applied magnetic field.} 
    
    \label{fig1} 
\end{figure}

We begin by characterizing the differential resistance in the $(I_1,I_2)$-plane of the three-terminal Josephson junction with the T-shaped geometry shown in Fig.~\ref{fig1}(a). 
Fig.~\ref{fig1}(b)-(d) show the differential resistances $dU_{13}/dI_1$, $dU_{12}/dI_1$ and $dU_{23}/dI_1$, as functions of the bias currents $I_1$ and  $I_2$ applied to the corresponding superconducting leads 1 and 2. Three distinct bias-current regimes can be identified, depending on how many JJs remain superconducting: {\it Low bias current}---
In this regime, all junctions are superconducting. The corresponding central area, colored dark blue in all three measurements, shows $dU_{ij}/dI_1=0$. It is enclosed by the CCC, a high resistive boundary characteristic of the transition from the superconducting to the resistive state. 
{\it Intermediate bias current}---At these currents, at most one junction remains superconducting. The differential resistance exhibits a six-lobed pattern in the $(I_1,I_2)$-plane, similar to that reported in Ref.~\cite{pankratova_multiterminal_2020}.
{\it Large bias current}---For bias currents outside the six-lobed pattern, all junctions are driven into the resistive regime and no additional features appear.

\subsection{$(I_1,I_2)$-plane}

We now analyze the extent of the superconducting regions in terms of the individual critical currents in the $(I_1,I_2)$-plane. 

\subsubsection*{Intermediate bias regime}

In this regime, each arm of the six-lobed pattern can be linked to an individual JJ$_{ij}$ by noting that the measured differential resistance remains in the superconducting regime $(dU_{ij}/dI_1=0)$ along the arm corresponding to that junction. 
Although this feature is barely visible in Fig.~\ref{fig1}(b) due to the small critical current $I_\text{c,13}$, it becomes more pronounced in Fig.~\ref{fig1}(c),(d), where  $dU_{12}/dI_1$ and $dU_{23}/dI_1$ exhibit broader superconducting arms.

The width and slope of each arm of the six-lobed pattern depends not only on the corresponding individual critical currents but also on the relative resistances of all junctions. To gain quantitative insight into their relationship, we use a multiterminal version of the RSJ model, which provides explicit expressions for the arm widths $\Delta I_1$ and $\Delta I_2$ in $I_1$ and $I_2$, as well as for the arm slopes as a function of $I_2$. The results are summarized in Table \ref{tab:interslopes}, with further details provided in App.~\ref{app.c}. Using the measured resistances of the individual junctions at 20\,mK, i.e.~$R_{13}=53.7\,\Omega$, $R_{12}=50.7\,\Omega$ and $R_{23}=35.9\,\Omega$ together with the expressions given in Table \ref{tab:interslopes}, we extract the individual critical current values: $I_\text{c,13}\approx 0.09\,\mu$A, $I_\text{c,12}\approx 0.39\,\mu$A and $I_\text{c,23}\approx 0.83\,\mu$A, from the measurements shown in Fig. \ref{fig1}(b)-(d).\\

\begin{table}[h]
    \centering
    \begin{tabular}{|c|c|c|c|}
    \hline
                  & $\Delta I_1$ & $\Delta I_2$ & slope\\
          \hline
        JJ$_{13}$ &  2 $I_\text{c,13}$ & $2(1+\frac{R_{12}}{R_{23}})I_\text{c,13}$ & -$\frac{R_{23}}{R_{12}+R_{23}}$ \\
        JJ$_{12}$  & $2(1+\frac{R_{23}}{R_{13}})I_\text{c,12}$ & 2 $(1+\frac{R_{13}}{R_{23}})I_\text{c,12}$ & $\frac{R_{23}}{R_{13}}$\\
        JJ$_{23}$& $2 I_\text{c,23}$ & 2 $(1+\frac{R_{12}}{R_{13}})I_\text{c,23}$ & -$(1+\frac{R_{12}}{R_{13}})$\\
         \hline
    \end{tabular}
    \caption{Widths and slopes of the three arms in the intermediate bias regime in relation to individual resistances $R_{ij}$ and critical currents $I_\text{c,ij}$, obtained from a multiterminal version of the RSJ model.}
    \label{tab:interslopes}
\end{table}

\subsubsection*{Low bias regime}
\label{lbr}
As shown in Fig. \ref{fig1}(b)-(d), the extent of the area enclosed by the CCC is not determined by a single critical current $I_{\text{c,}i}$ like in isolated JJs. Instead, it results from a linear combination of the individual critical currents, i.e.~$I_{\text{c,}ij}^\text{comb}=\sum_i \alpha_i I_{\text{c,}i}$. 
For the geometry shown in Fig. \ref{fig1}(a) with lead 3 grounded and $I_2=0$, the bias current $I_1$ flows through two paths: One containing a single JJ, $\text{JJ}_{13}$, and the other containing two JJs in series, i.e.~$\text{JJ}_{12}$ and $\text{JJ}_{23}$. As in any parallel circuit, the voltage across each path is equal. A finite voltage develops only when the applied bias current drives both arms into the resistive regime. As a result, the combined critical current is given by

\begin{align}
\label{eq:combcc}
I^{\text{comb}}_\text{c,13}(I_1,I_2=0)\sim I_\text{c,13}+ \text{min}\{I_\text{c,12},I_\text{c,23}\},
\end{align}
with $I_{\text{c},ij}$, the individual critical current components. 

Similarly, if  a current bias $I_2$ is applied and $I_1=0$, the parallel circuit consists of $\text{JJ}_{23}$ on one path and $\text{JJ}_{12}$ and $\text{JJ}_{13}$ along the other. In this case, the effective critical current is approximately given by 
\begin{align}
    I^{\text{comb}}_\text{c,13}(I_1=0,I_2)\sim I_\text{c,23}+\text{min}\{I_\text{c,13},I_\text{c,12}\}, \label{eq:iccomb2}
\end{align}
provided that $I_\text{c,23}>I_\text{c,13}$~\footnote{Note that the condition $I_\text{c,23}>I_\text{c,13}$ comes from the fact that the voltage drop taking place in the JJ$_{13}$ shares the arm with JJ$_{12}$. If $I_\text{c,13}>I_\text{c,23}$, then, $I^{\text{comb}}_\text{c,13}(I_1=0,I_2)$ is given by the intermediate bias current regime, $(1+R_{12}/R_{23})I_\text{c,13}$, see more details in SM.}.
These estimations are confirmed using an adapted version of the RSJ model, which qualitatively reproduces the behavior observed in the measurements as a function of both bias current and applied magnetic field (see App.~\ref{app.c} for details). 

From Fig. \ref{fig1}(b), we extract $I^{\text{comb}}_\text{c,13}(I_1,I_2=0)\approx 0.43-0.48\,\mu$A and $I^{\text{comb}}_\text{c,13}(I_1=0,I_2)\approx 1.36-1.5\,\mu$A. 
Using the extracted critical current values of the individual JJs together with Eqs.~\eqref{eq:combcc} and~\eqref{eq:iccomb2}, we then obtain $I^{\text{comb}}_\text{c,13}(I_1,I_2=0)\approx 0.48\,\mu$A and $I^{\text{comb}}_\text{c,13}(I_1=0,I_2)\approx 0.93\,\mu$A. While $I^{\text{comb}}_\text{c,13}(I_1,I_2=0)$ matches the value observed in Fig.~\ref{fig1}(b) quite accurately, $I^{\text{comb}}_\text{c,13}(I_1=0,I_2)$, shows a significant deviation. This discrepancy may arise from capacitance effects, which can effectively increase the critical current, and possibly from Joule heating effects, which can contribute to the underestimation of the individual critical currents.

\subsection{Magnetic field dependence}

\begin{figure}[t]
\centering
    \includegraphics[width=0.7\textwidth, keepaspectratio] {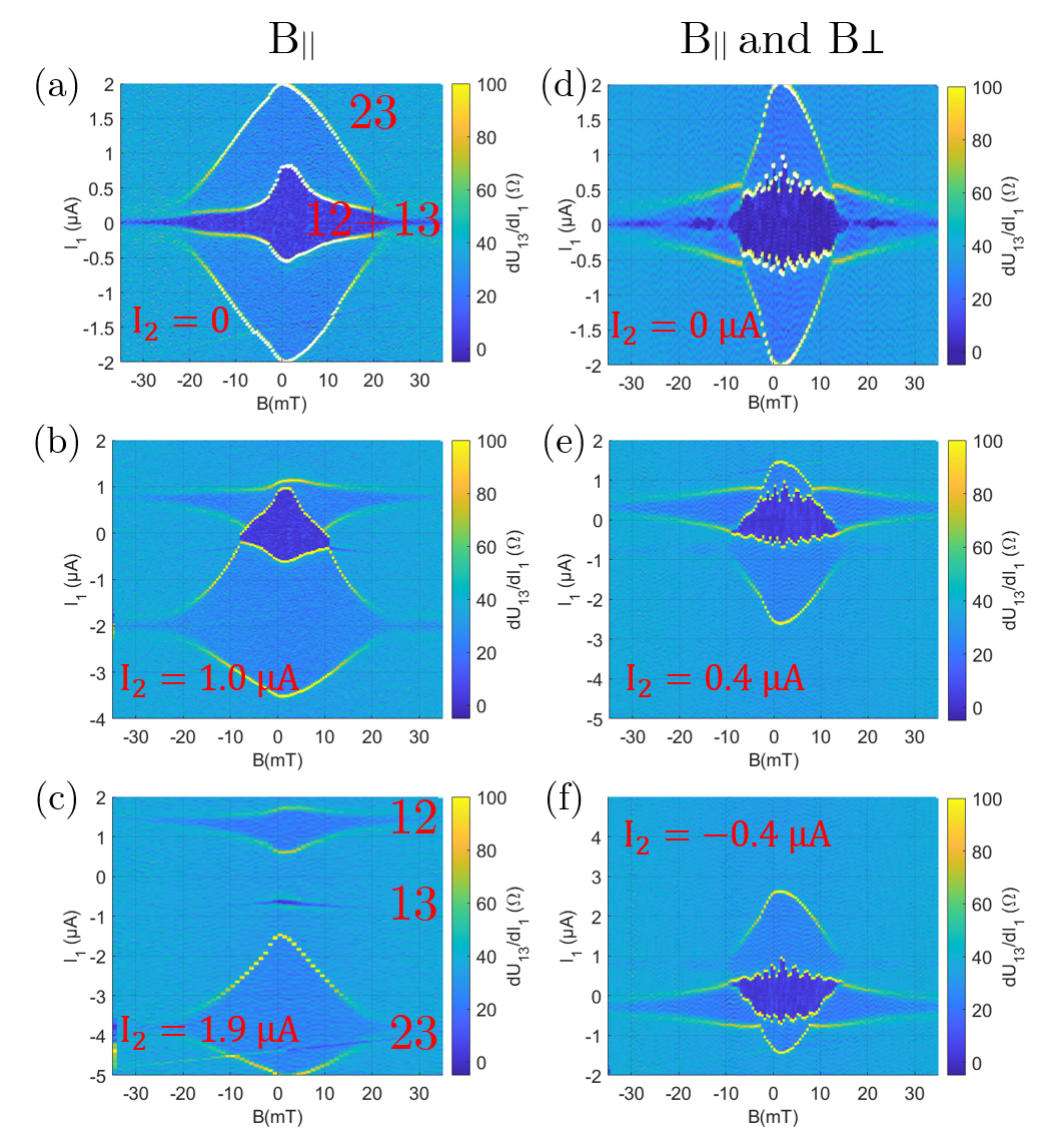}
    \caption{
    \textbf{Influence of a bias current and applied magnetic field on the T-MTJJ}.
    (a)-(c): Differential resistance $dU_{13}/dI_1$ as a function of $I_1$ and applied in-plane magnetic field $B$ for $I_2=0,~1,~1.9\,\mu$A, respectively.    
    (d)-(f) Differential resistance $dU_{13}/dI_1$ as a function of $I_1$ and applied magnetic field $B$, with the sample slightly rotated giving rise an out-of plane component, for $I_2=0,0.4,-0.4\,\mu$A. 
    }\label{fig2} 
\end{figure}    

We continue our analysis by studying the influence of an external magnetic field on the critical current extracted from the differential resistance measurements $dU_{13}/dI_1$. To this end, we set the bias current to constant values of $I_2=0,1~\text{and}~1.9\,\mu$A, as indicated by the red dashed lines in Fig.~\ref{fig1}(b) and measure $dU_{13}/d I_1$ as a function of $B$ and $I_1$, as shown in Fig.~\ref{fig2} ~\footnote{Note that the critical currents measured in the $(I_1-B)$-plane are larger than those shown in Fig \ref{fig1}. This is due to a residual, non zero, magnetic field present during the differential resistance measurements in the $(I_1,I_2)$-plane measurements. An additional $I_1,I_2$ figure with closer values is shown in App.~\ref{app.a}.}.

When the applied magnetic field is in-plane (Fig.~\ref{fig2}(a)-(c)), we observe an exponential suppression of the critical current, resembling the first lobe of a Fraunhofer pattern. 
Introducing a finite bias current $I_2$ shifts the superconducting region (dark blue) away from the center $(I_1=0)$, resulting in an asymmetric lobe shape, bounded by the overlap between the envelopes of $\text{JJ}_{23}$ and $\text{JJ}_{12}$ as indicated in Fig.~\ref{fig2}. Note the asymmetry of the dark blue areas with respect to $I_1=0$ at finite $I_2$, which will be important below when discussing the superconducting diode effect.

We now rotate the sample so that the applied magnetic field acquires an out-of-plane component $B_\perp$. Under these conditions, the CCC exhibits an oscillatory pattern --resembling a SQUID-like interference pattern-- superimposed on the exponential suppression, see~\ref{fig2}(d)-(f). The out-of-plane component corresponds to a tilt of $\approx 0.986^{\circ}$ from the in-plane orientation. The resulting periodicity of the pattern matches the value expected for the MTJJ when the current is transported along the edge of the superconducting leads. A more detailed analysis is provided in the App.~\ref{app.b}. 
These oscillations reveal the two-path interferometer circuit inherent to this geometry: within the CCC, all three junctions remain in the superconducting regime and the Andreev bound states extend coherently across all three junctions. In this scenario, a particle entering through lead 1 propagates to lead 3 via two paths: JJ$_{13}$ and JJ$_{12}$-JJ$_{23}$. Along each path, the particle acquires a different Peierls phase, giving rise to the observed oscillations.
A finite $I_2$ confines the CCC region to the overlap of the $\text{JJ}_{23}$ and $\text{JJ}_{12}$ envelopes, causing the SQUID-like pattern to become distorted with an asymmetry determined by the sign of $I_2$, see Fig.~\ref{fig2}(e) and~(f).

\subsection{Magnetic field in the ($I_1$,$I_2$)-Plane}

Having discussed the SQUID-like pattern, we now turn to the impact of a perpendicular magnetic field on the CCC in the $(I_1,I_2)$-plane. To illustrate this effect, we use the X-shaped junction, see Fig.~\ref{fig3}(a), where this is more pronounced than in the T-junction~\footnote{Note that the relative visibility of the CCC deformation depends on the smallest critical current. In the case of the T-junction, the smallest $I_\text{c,13}=0.09\,\mu$A, so that the deformations are barely visible.}. For this device the contacts only worked partially, meaning that only 3 point measurements were possible leading to a finite resistance of $\approx2400~\Omega$ instead of real $0~\Omega$. Fig.~\ref{fig3}(b) shows measurements of the differential resistance $dU_{13}/dI_1$ as a function of $I_1$ and the applied magnetic field $B$. As in the T-junction, a SQUID-like pattern appears in the low bias regime $(I_2=0)$~\footnote{Note that the SQUID-pattern begins from a minimum instead of a maximum. This can be caused by the presence of a trapped flux, or the magnetochiral effect. However, we did not investigate further its source.}.
Interestingly, the CCC is periodically deformed in a characteristic manner across the $(I_1,I_2)$-plane, as shown in Fig.~\ref{fig3}(c). It shows the differential resistance $dU_{13}/dI_1$ in the $(I_1,I_2)$-plane for four values of the applied magnetic field, $B_0$, $B_1$, $B_2$, $B_3$, indicated by the colored arrows in Fig.~\ref{fig3}(b). For $B=B_1$ and $B_3$ ($B_0$ and $B_2$), the initially symmetric CCC lobe becomes asymmetrically (symmetrically) in the central region. The sequence of CCC shapes observed in Fig.~\ref{fig3}(c) reappears periodically at subsequent SQUID-like lobes shown in Fig.~\ref{fig3}(b).
To gain insight into the influence of the magnetic field on the CCC, we employ the RSJ-model and find  similar behavior for magnetic fluxes $\Phi/\Phi_0=0, 1/2, 1,$ and 3/2, as shown in Fig.~\ref{fig3}(d). In this model, the magnetic field enters by shifting the superconducting phase differences as~\cite{tinkham2004}: 

\begin{figure}[]
    \includegraphics[width=1\textwidth, keepaspectratio] {
    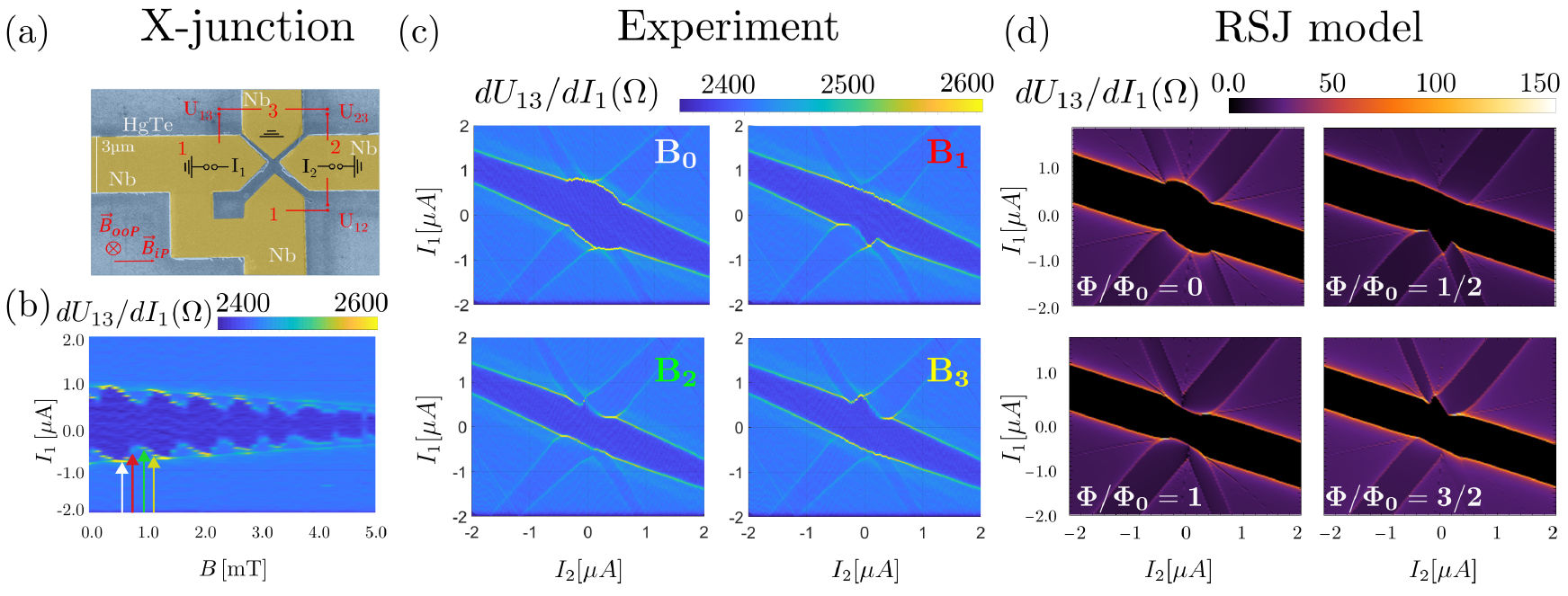}
    \caption{
    \textbf{$dU_{13}/dI_1$ vs $I_1$ and $I_2$ for different applied magnetic fields in the X-junction: Experimental and theoretical results.} 
    (a): False-colored SEM schematic of the four terminal JJ. Yellow indicates the \text{Nb} leads, with two contacts shorted. 
    (b): Differential resistance $dU_{13}/dI_1$ as a function of the applied magnetic field $B$, showing in a SQUID-like pattern. (c): $dU_{13}/dI_1$ as a function of $I_1$ and $I_2$ for four different applied magnetic fields $B_0,B_1,B_2,B_3$, indicated by colored arrows in Fig. \ref{fig:RSJ_I1_B} (b).
    (d): Corresponding theoretical results, from the RSJ model, using $I_\text{c,13}=0.55\,\mu$A, $I_\text{c,12}=0.27\,\mu$A, $I_\text{c,23}=0.2\,\mu$A and $R_{13}=25.0\,\Omega$, $R_{12}=53.7\,\Omega$, $R_{23}=40.9\,\Omega$. 
    }\label{fig3} 
\end{figure}   

\begin{align}
\pi\frac{\Phi}{\Phi_0}=-\phi_{13}+\phi_{12}+\phi_{23},
\end{align}
with $\phi_{ij}$ the superconducting phase difference between the superconducting leads $i$ and $j$, the magnetic flux $\Phi=B_\perp \mathcal{S}$, the flux quanta $\Phi_0=h/2e$ and $\mathcal{S}$ the surface delimited by the area enclosed by the superconducting electrodes, see more details in App.~\ref{app.c}.

The magnetic flux $\Phi/\Phi_0$ modifies the functional form of one of the supercurrents, i.e.~$I_{\text{c},23}\sin(\phi_{13}-\phi_{12})\rightarrow I_{\text{c},23}\sin(\phi_{13}-\phi_{12}+\pi \Phi/\Phi_0)$. 
Accordingly, the sign of this supercurrent is effectively reversed when $\Phi/\Phi_0$ changes from $0$ to $1$. This reduces the combined critical current to $ I^{\text{comb}}_\text{c,13}(I_1,I_2=0)\sim |I_\text{c,13}- \text{Min}\{I_\text{c,12},I_\text{c,23}\}|$. 
Consequently, the difference between the maxima and the minima of the SQUID pattern becomes,
\begin{align}
\Delta I^{\text{comb}}_\text{c,13}(I_1,I_2=0)\sim 2\min\{I_\text{c,13},  I_\text{c,12}, I_\text{c,23}\}.
\end{align}

For $\Phi/\Phi_0=1/2$ and $3/2$ the supercurrent term that originally follows a sine transforms into a cosine. 
This change in the functional form of one of the supercurrents modifies the CCC in a non-trivial way, 
giving rise to an asymmetric pattern in the $(I_1,I_2)$-plane:
a $(-)\cos$ generates a lobe at $I_1<0$ ($I_1>0$) and a flat CCC at $I_1>0$ ($I_1<0$). 

\subsection{Superconducting diode effect}

The superconducting diode effect ideally manifests as a non-resistive current when the junction is biased in one direction while becoming resistive when the bias is applied in the opposite direction. This asymmetry requires the breaking of both time-reversal and inversion symmetry \cite{baumgartner_supercurrent_2022,Fracassi2024,reinhardt_link_2024, Lombardi2025, costa_unconventional_2025}. In MTJJs, this effect can even appear at zero magnetic field, where time-reversal symmetry is effectively broken by applying a bias current through one of the superconducting electrodes~\cite{chiles_nonreciprocal_2023}. Although the sample design in Ref.~\cite{chiles_nonreciprocal_2023} differs - featuring a superconducting island connecting the three junctions - the underlying physics, namely phase differences between the junctions that deviate from 0 or integer multiples of $\pi$, operates in the same way.

\begin{figure}[t]
\centering
    \includegraphics[width=0.7\textwidth, keepaspectratio] {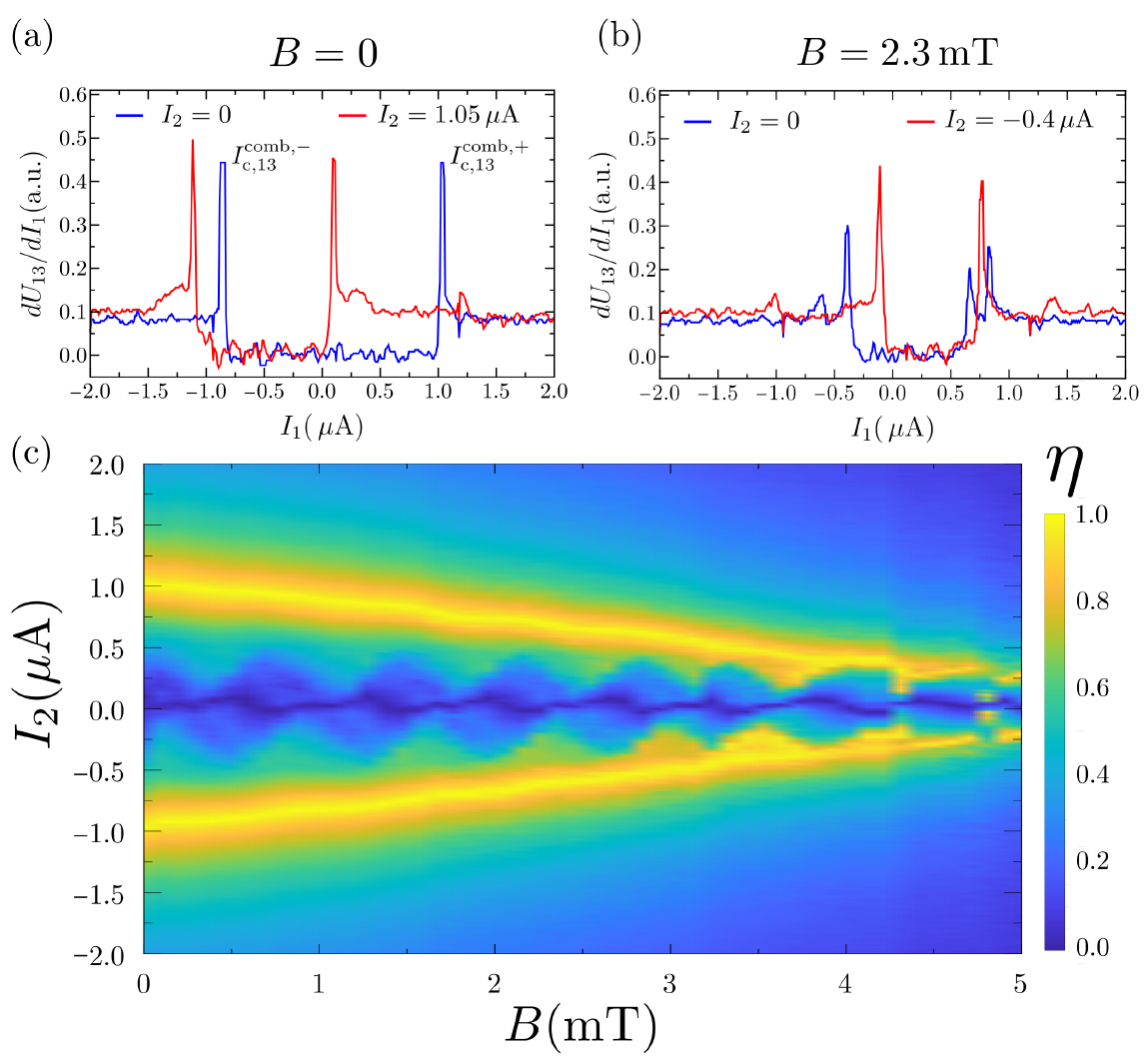}
    \caption{
    \textbf{Superconducting diode effect in the X-junction}---  
(a)-(b): Differential resistance $dR_{13}/dI_1$ as a function of the bias current $I_1$, for $B=0$ (a) and $B=2.3\,\text{mT}$ (b). Blue curves correspond to $I_2=0$ and red curves to $I_2\neq 0$.
 (c): Diode efficiency $\eta$ extracted from the experimental results for the three-terminal Josephson junction with X-shaped geometry shown in Fig.~\ref{fig3}. The efficiency is plotted as a function of the bias current $I_2$ and magnetic field $B$.
    }\label{fig4} 
\end{figure}

A finite $I_2$ shifts $I_{\text{c},13}^{\text{comb}}$ asymmetrically with respect to the sign of $I_1$, resulting in two different critical currents, $I_{\text{c},13}^{\text{comb},\pm}$, for $I_1>0$ and $I_1<0$, see Fig.~\ref{fig4}. For $I_2>0$, the critical current $I_{\text{c},13}^{\text{comb},+}$ shifts towards 0, while $I_{\text{c},13}^{\text{comb},-}$ is shifted towards larger values; the opposite occurs for $I_2<0$.

We can see the diode effect more clearly in Fig.~\ref{fig4}(a), where we plot $dR_{13}/d I_1$ as a function of $I_1$ at $B=0$, $I_2=0$ (blue) and $I_2=1.05\,\mu$A (red). For $I_2=0$, the two resistive peaks are symmetrically distributed around $I_1$, with $I_{\text{c},13}^{\text{comb},\pm}\approx \pm 1.0\,\mu$A. When a bias current of $I_2\approx 1.0\,\mu$A is applied, $I_{\text{c},13}^{\text{comb},+}$ shifts towards 0, while $I_{\text{c},13}^{\text{comb},-}$ remains finite. At finite magnetic fields, $B\neq 0$, the superconducting diode effect appears even for $I_2=0$, as time-reversal symmetry is explicitly broken, see the blue curve in Fig.~\ref{fig4}(b). Furthermore, for finite magnetic fields, we find $I_{\text{c},13}^{\text{comb},-}\approx 0$ in the low bias regime, with $I_2\approx 0.4\,\mu$A. This behavior can be attributed to the overall reduction of the individual critical currents $I_{\text{c},ij}$ at finite magnetic fields and to the distortion of the CCC, shown in Fig.~\ref{fig3}(c).

To quantify the superconducting diode effect, we define the diode efficiency $\eta$ as
\begin{align}
    \eta=\left| \frac{I_\text{c,13}^{\text{comb},+}-|I_\text{c,13}^{\text{comb},-}|}{I_\text{c,13}^{\text{comb},+}+|I_\text{c,13}^{\text{comb},-}|}\right |,
\end{align}
which is shown in Fig.~\ref{fig4}(c) as a function of the bias current $I_2$ and magnetic field $B$. As expected from our previous discussion, for $B=0$, the efficiency $\eta$ increases as a function of $I_2$ reaching $\eta\sim 1$ for $I_2\approx \pm 1.0\,\mu$A. For finite $B$, the bias current $I_2$ needed to obtain $\eta=1$ is reduced due to the overall suppression of the individual critical currents with increasing $B$. 

In the low-bias regime $|I_2|\lesssim 0.5\,\mu$A, $\eta$ exhibits an oscillatory pattern reflecting the same coherent effects that give rise to the SQUID-like interference and the asymmetric deformation of the CCC. In this regime, large efficiency regions appear periodically, with $\eta\sim 0.7- 0.8$ for $B\lesssim 3\,$mT.

In general, enhancing $\eta$ by applying a large bias current suppresses coherence effects between the three junctions, as evidenced by the absence of oscillatory patterns in the top and bottom yellow stripes of Fig.~\ref{fig4}(c). At such large bias current, the superconducting diode effect is no longer observed in other differential resistances $dR_{ij}/dI_1$, with $ij\neq 13$. In contrast, in the low bias regime, regions with large diode efficiencies appear in all three differential resistances $dR_{ij}/dI_1$, as coherence effects between the three junctions are preserved. A similar scenario was discussed within the framework of scattering theory in Ref.~\cite{correa_theory_2024}.


\section{Conclusions and outlook} \label{sec:conclusions}

In this work, we investigated two fully coupled three-terminal Josephson junctions based on \text{Nb} and \text{HgTe}, realized in distinct T-shaped and X-shaped geometries. Our study focused on the properties of the critical current contour (CCC), which revealed rich and tunable behavior arising from the interplay of geometry, current bias, and magnetic fields.

The first key finding is that the CCC extends beyond the simple sum of the individual critical currents. For instance, at $ I_2 = 0 $, we observe a combined critical current approximately given by  
\begin{align}
I_\text{c,13}^{\text{comb}}(I_1,I_2=0)\approx I_{\text{c},13}+\min\{I_{\text{c},12}, I_{\text{c},23}\},
\end{align}
highlighting the nontrivial coupling between the terminals and setting the stage for magnetic field tunability.

By probing the CCC response to magnetic fields, we find that in-plane fields cause an exponential suppression of the critical current, whereas out-of-plane fields generate a pronounced SQUID-like interference pattern. The amplitude of this pattern scales as  
\begin{align}
\Delta I^{\text{comb}}_\text{c,13}(I_1, I_2=0) \sim 2\min\{I_\text{c,13},  I_\text{c,12}, I_\text{c,23}\},
\end{align}
highlighting the cooperative role of adjacent junctions.

Our second finding is that the addition of a bias current introduces a crossover from this SQUID-like to a Fraunhofer-like pattern, providing an additional degree of control over the superconducting state. Furthermore, a out-of-plane magnetic field produces periodic, flux-dependent deformations of the CCC in the $(I_1, I_2)$-plane. These distortions originate from the gauge-invariant phase relation  
\begin{align}
\pi\frac{\Phi}{\Phi_0} = -\phi_{13} + \phi_{12} + \phi_{23},
\end{align}
and appear as symmetric or asymmetric, depending on whether $\Phi/\Phi_0$ is integer or half-integer. This is another novel mechanism that these asymmetric deformations enhance the superconducting diode effect in the low-field, low-bias regime, achieving diode efficiencies as high as $\eta \approx 0.8$.

To conclude, our quantitative understanding of how bias currents and magnetic fields determine the superconducting regime of a three-terminal Josephson junction enables accurate prediction of its behavior from the intrinsic gaps and resistances of the individual junctions. This predictive capability provides a powerful foundation for the design of multiterminal superconducting circuits. In quantum technologies, it can improve device reproducibility, allow engineered Josephson couplings, and enhance the stability of superconducting qubits and hybrid circuits. At the same time, the sensitivity of three-terminal junctions to multiple phase differences makes them promising candidates for phase-sensitive metrology and precision magnetometry. Thus, despite the absence of topological features in our measurements, the framework presented here bridges fundamental superconducting physics with practical device engineering, paving the way toward controlled, tunable, and scalable superconducting quantum and sensing architectures.

\section*{Acknowledgements}
We acknowledge stimulating discussions with M. Stehno.

\paragraph{Funding information}
This work was funded by the European Research Council under the European Union’s Horizon 2020 research and
innovation program (Grant Agreement No.~787515, 253 ProMotion). We also acknowledge support by the Deutsche Forschungsgemeinschaft (DFG, German Research Foundation) within Project-ID 314695032 -- SFB 1277 (projects A07, A08, B08) and through the Elitenetzwerk Bayern Doktorandenkolleg "Topological Insulators". F.D. is grateful for funding support from the Deutsche Forschungsgemeinschaft (DFG, German Research Foundation) under Germany’s Excellence Strategy through the W\"urzburg-Dresden Cluster of Excellence on Complexity and Topology in Quantum Matter ct.qmat (EXC 2147, Project ID
390858490) as well as through the Collaborative Research
Center SFB 1170 ToCoTronics (Project ID 258499086). G.P. acknowledges support from Spain’s MINECO through Grant No. PID2023-149072NB-I00 and by the CSIC Research Platform PTI-001.


\bibliography{library_multiterminal_28_07_25.bib}

\newpage
\begin{appendix}

\clearpage

\section{Fabrication and measurement setup}

\subsection{Fabrication}

The devices investigated in this work are based on the topological insulator \text{HgTe}, grown in a lattice-matched configuration between layers of \text{Cd}$_{0.7}$Hg$_{0.3}$\text{Te} and \text{CdTe} on a \text{GaAs} substrate (see Fig\ref{appa1} (a)). Details of the growth by molecular beam epitaxy can be found in \cite{dantscher_cyclotron-resonance-assisted_2015}.

\subsubsection*{Mesa etching}
\begin{figure}[h]
\centering
    \includegraphics[width=0.24\textwidth, keepaspectratio] {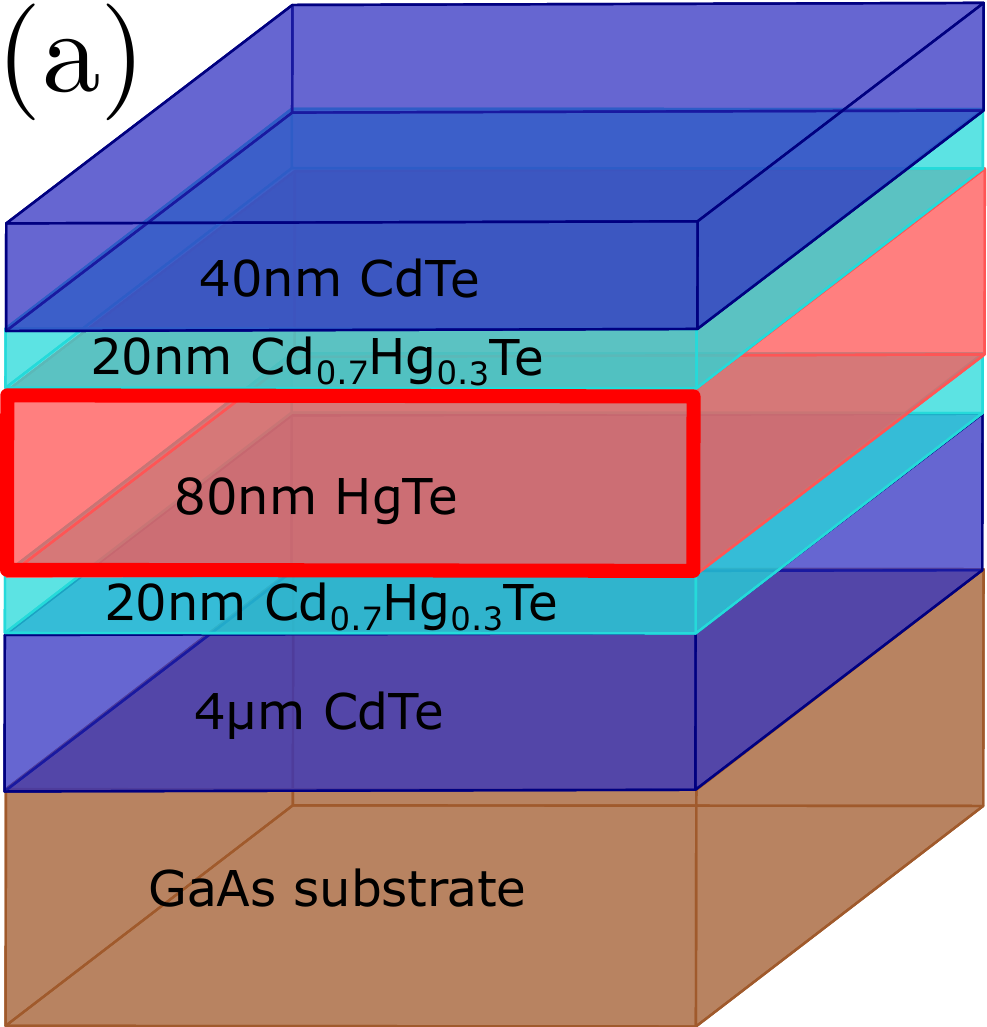}
    \includegraphics[width=0.5\textwidth, keepaspectratio] {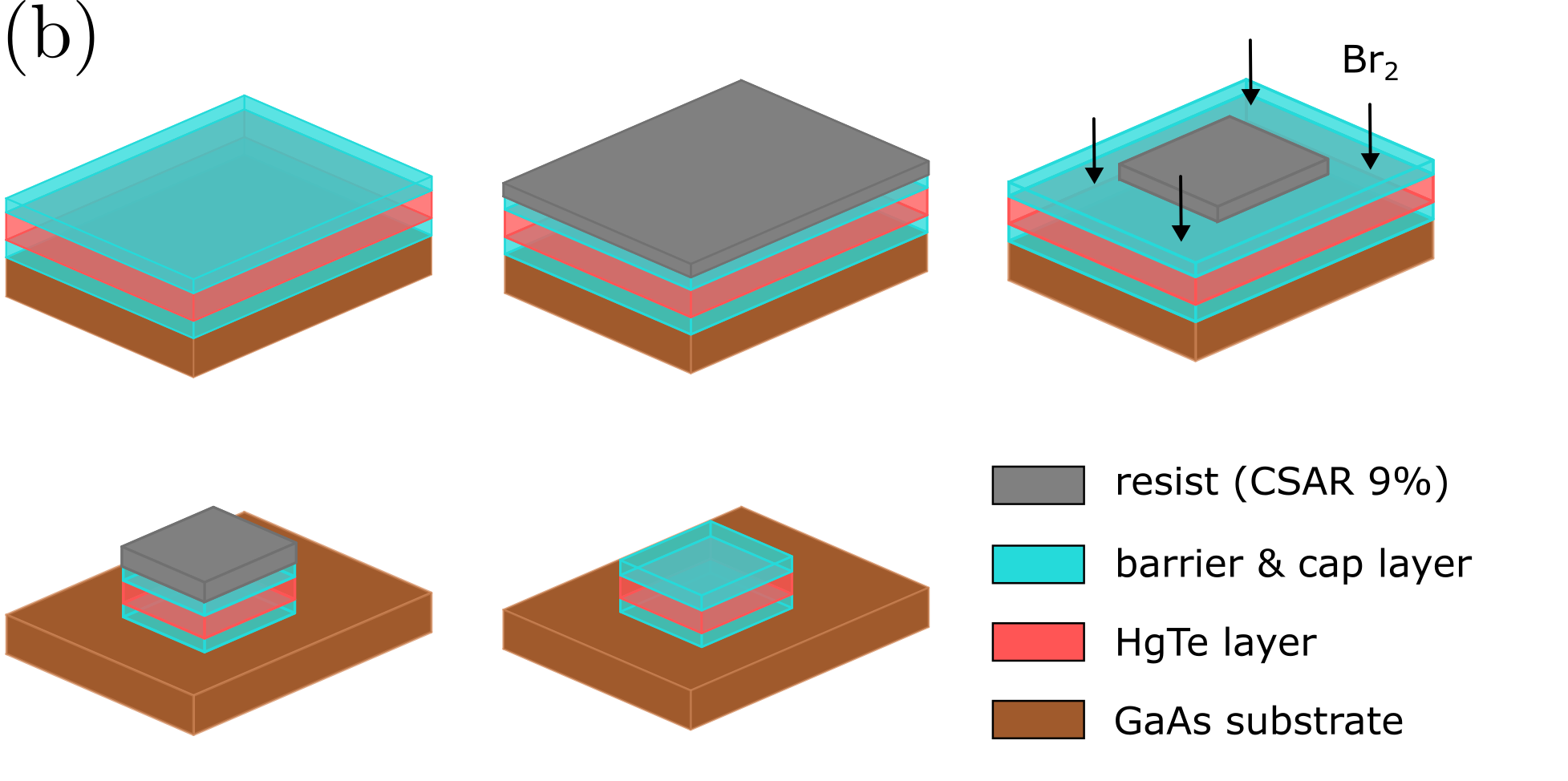}
    \caption{
    \textbf{Schematic of the wafer stack and the mesa fabrication steps} (a) shows the used wafer stack, in which the \text{HgTe} is sandwiched between two layer of \text{CdHgTe}. (b) depicts the fabrication steps to for the mesa. Fabrication for the MTJJ is done in the same way.
    }\label{appa1} 
\end{figure}

The mesa structure was defined by electron beam lithography (EBL) using  CSAR 9\,\% resist, spin-coated and baked at 80\,°C for 25\,min. After exposure and development in AR600-546, a $15\,\mu \text{m} \times 15\,\mu \text{m}$ mesa was etched in a Br-based wet etching solution (0.1\,ml \text{Br}$_{2}$, 100\,ml ethylene glycol, 25\,ml \text{H}$_2$\text{O} at $0$ °C for  5\,min, followed by a water rinse and  \text{N}$_{2}$ blow-dry. The resist was subsequently removed with AR600-71 (see Fig. \ref{appa1}(b)).\\

\subsubsection*{Junction definition}
The multiterminal Josephson junctions (MTJJs) were patterned by EBL using the same resist and development process. Junction lengths between between $100$ and $300$ \,nm and widths between $0.5$ and $5\,\mu \text{m}$ were defined and etched in a diluted Br-based solution (\text{Br}$_{2}$), $100$ ml ethylene glycol and $25$ ml \text{H}$_{2}$\text{O}) at$0$ °C for $90$ s, followed by rinsing and drying. The sample was then transferred into an ultra-high vacuum (UHV) chamber, where the exposed \text{HgTe} surface was cleaned by \text{Ar}$^+$ pre-sputtering for $10$ s. A \text{Ti} adhesion layer of $4-5$ nm was deposited by electron-gun evaporation, followed by $80$ nm of \text{Nb} grown by \text{Ar} DC sputtering, and capped with $5$ nm of \text{Au} to prevent oxidation. Lift-off was performed in AR600-71.\\

\subsubsection*{Contact definition}
Contact leads connecting the bonding pads to the MTJJs were defined by EBL, developed in AR600-546, and cleaned by \text{Ar}$^+$ pre-sputtering for  $10$s. A Ti/Au metallization ($4-5$ nm / $100$ nm) was deposited by electron-gun evaporation, followed by lift-off in AR600-71.\\

\subsubsection*{Device design and selection}
The junction geometry was chosen in accordance with Ref. \cite{pankratova_multiterminal_2020}: the Y-shaped structure is expected to yield similar critical currents across all three contacts, while the T-shaped structure allows independent alignment of the magnetic field parallel and perpendicular to the individual junctions. As these were the first MTJJs fabricated on \text{HgTe} using this process, a broad range of junction dimensions was realized - superconductor widths from 500 \,nm to 5 \,$\mu$m and lengths from $100$ to $300$ \,nm - chosen in accordance with those reported in previous works \cite{Fischer2022_4pi, Himmler2023_supercurrent_interference} and to ensure sufficient junction area for meaningful magnetic field measurements. The junction dimensions were not specifically optimized with respect to the Nb coherence length. Following SEM inspection, devices free of visible structural defects were selected for low-temperature characterization. Previous investigations of the\text{HgTe-Nb} interface report that the Fermi energy is typically located near the top of the valence band, with interface transparencies of $D\approx0.43-0.70$ \cite{Fischer2022_4pi, Himmler2023_supercurrent_interference}.  Further properties of the HgTe-Nb interface, typical for the used material can be found here \cite{maier_induced_nodate, sochnikov_nonsinusoidal_2015, wiedenmann_transport_2017,kozlov_transport_2014}.

\begin{figure}[h]
\centering
    \includegraphics[width=0.7\textwidth, keepaspectratio] {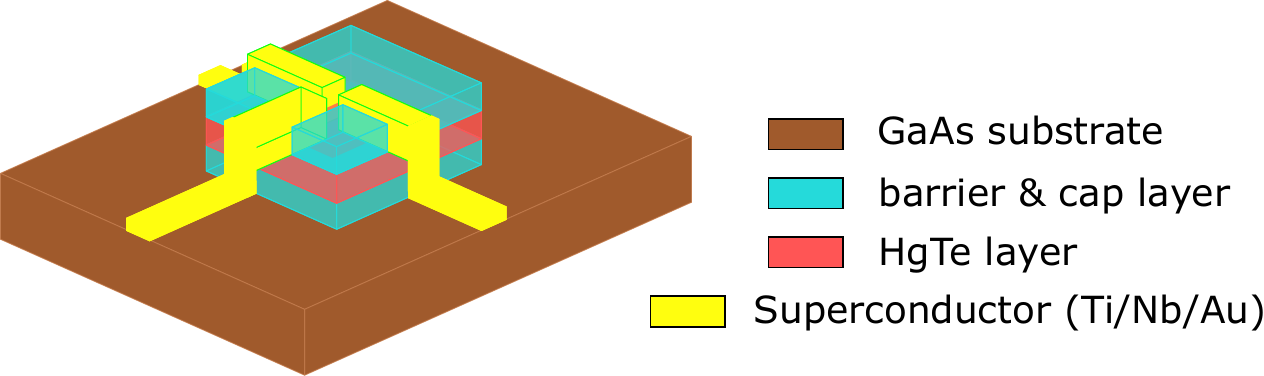}
    \caption{
    \textbf{Schematic of the finished sample} Shown are the superconducting contacts of the MTJJ and the etched mesa on top of the \text{GaAs} substrate. The further connection from the junction to the bond pads is left out for a better visibility of the MTJJ. 
    }\label{appa2} 
\end{figure}

\subsection{Measurement setup}
\begin{figure}[h]
\centering
    \includegraphics[width=0.7\textwidth, keepaspectratio] {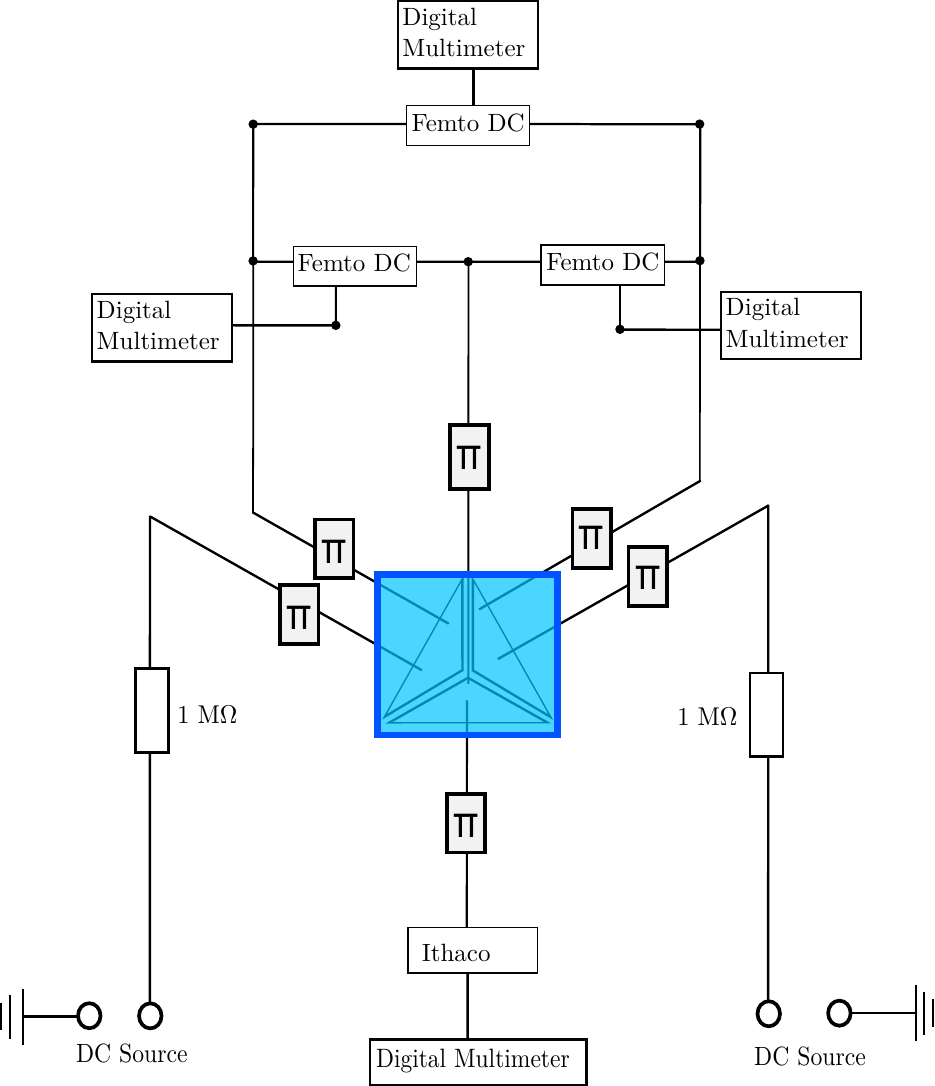}
    \caption{
    \textbf{Measurement setup sketch} The used measurement setup used is depicted with all the devices used at room temperature. The blue area marks the cryostat, its wiring and the device inside.
    }\label{appa3} 
\end{figure}

Both samples were measured in an \textit{Oxford Instruments Kelvinox TLM} dilution refrigerator at a base temperature of approximately $20\,\text{mK}$. The cryostat is equipped with a superconducting magnet and a rotatable sample holder, allowing magnetic fields to be applied at arbitrary angles relative to the sample. To reduce Johnson–Nyquist noise, all lines inside the cryostat are equipped with RC low-pass filters (cutoff frequency $10$ kHz) and attenuators providing more than $100$ dB of attenuation in the frequency range $150$ MHz–$10$ GHz, and more than $50$ dB above $10$ GHz.\\
At room temperature, the DC signals were applied and recorded using a \textit{Nanonis Tramea} system. Each line was additionally equipped with a $1.5$ nF $\pi$-filter (\textit{Tusonix 4201-000LF}) to suppress high-frequency interference. A quasi-four-point measurement scheme was employed, as illustrated in Fig. \ref{appa3}: two terminals supplied current to the sample through a series resistor ($1$ M$\Omega$), while a third terminal served as the current drain. The resulting current signal was amplified by a current-to-voltage converter (\textit{Ithaco 1211 current Preamplifier}) and subsequently recorded by the \textit{Nanonis Tramea} system. Each superconducting terminal was additionally connected to a dedicated voltage line, enabling the voltage drop between any pair of contacts to be measured independently. The voltage signal was first amplified using a \textit{Femto DLPVA-100-F-D} voltage preamplifier and then recorded by the \textit{Nanonis Tramea} system acting as a DC voltmeter. In this configuration, the sole low-resistance path to ground runs through the drain terminal via the current-to-voltage converter, ensuring that the entire current sourced into the device exits through the designated drain contact.

\clearpage
\section{Correlating the junctions to the areas in the plots}
\label{app.a}
\begin{figure}[h]
\centering
\includegraphics*[width=1\linewidth]{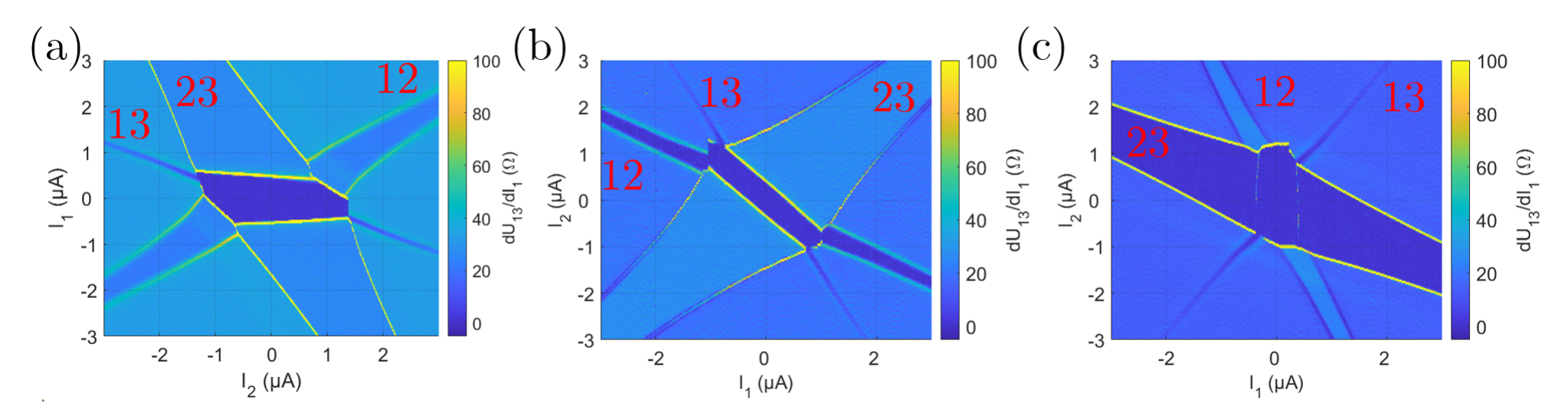} 
\caption{\textbf{Results of permuting the current contacts.} (a) Shows the same plot as in Figure \ref{fig1} (b), with all contacts connected as introduced before. In (b), the current contacts are permutated, so that $I_1$ is applied at contact 2 and $I_2$ at contact 3. This has an impact on the CCC and the junction in the zero resistance state. For (c) the current contacts are shifted once more from $2\rightarrow3, 3\rightarrow1$. } \label{app1}
\end{figure}

This paragraph introduces a second experimental method to distinguish the individual junctions in a MTJJ. The idea is that only the junction on the single side, directly between drain and sweeping current contact, always shows a zero resistance state, independent of the voltage used to calculate the differential resistance. For this purpose, we permute the current contacts in our setup, while keeping the voltage contacts fixed. In the standard configuration, introduced in Figure \ref{fig1} (a), $I_1$ is connected to contact 1 and $I_2$ to contact 2. For the first permutation, $I_1$ is applied to contact 2 and $I_2$ to contact 3, leading to $JJ_{12}$ on the single side. The result is shown in Figure \ref{app1} (b), where $JJ_{12}$ shows a zero resistance state in addition to $JJ_{13}$. For the second permutation $I_1$ is applied at contact 3 and $I_2$ is applied at contact 2. Therefore, $JJ_{13}$ is on the single side and hence shows a zero resistance state. From this we can correlate the physical junctions $JJ_{12},JJ_{13}$ and $JJ_{23}$ to the zero-resistance arms in the plots. Another possible interpretation is to examine the incline of each arm. Only one of them ever has a positive incline: the one between the two current contacts. Therefore, this junction can easily be spotted during the permutation.\\

\section{SQUID pattern correlation to the physical area}
\label{app.b}
\begin{figure}
   \includegraphics[width=1\textwidth]{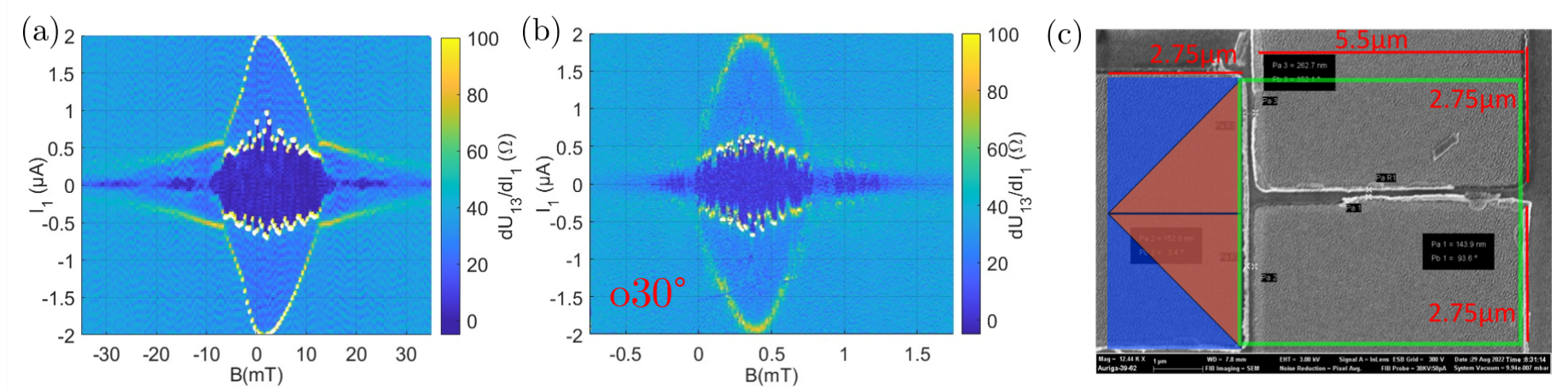}
	\caption[Area calculation for the SQUID.]{\begin{footnotesize}\textbf{Area calculation for the SQUID.} (b) Shows the $I_1-B$ plot with the highest out-of plane angle, where the oscillations can be clearly distinguished. (c)This SEM picture shows the dimensions of the areas corresponding to the SQUID oscillations. The green square represents the largest possible ring of the SQUID, within which the current could flow. The orange triangles mark exemplary areas for the flux-focusing effect and the read lines display their length and width. The flux focusing for the top and bottom lead is not shown here due to image size constraints.\end{footnotesize}}
	\label{Area_SQUID}
\end{figure}

The following paragraph focuses on the analysis of the smaller critical current and its oscillations. For that we present data in Figure \ref{Area_SQUID} (b), obtained by rotating the sample by $30^{\circ}$ into the plane of the sample, relative to the fully out-of-plane direction.
For larger angles, the difference between successive oscillations becomes smaller and can no longer be reliably determined from the graphs due to the limited step size in the used measurement setup. From the plot, we extract a field difference of $\Delta B = 0.05\,\text{mT}$. Using the relation $A=\frac{\phi_0}{\Delta B}$, the corresponding effective area is approximately $41.4\,\mu\text{m}^2$.To take into account that this area represents the projection of the actual area along the direction of the magnetic field, the corresponding area on the device surface is given by $A(0^{\circ})=\frac{A}{cos(30^{\circ})}\approx 48\,\mu\text{m}$ 

Since no clear short links between the superconducting leads are visible in the SEM picture (Fig. \ref{Area_SQUID} (c) ), we can only give a maximum estimation for the area. Therefore, we assume the short link to be located at the edge of the contacts (see the green square in Fig. \ref{Area_SQUID} (c) ). Calculating the area combined with the flux focusing effects (partially marked in orange) according to eq. \ref{SQUID_ff}, matches the expected area for an fully out-of-plane magnetic field.

\begin{align}
5.5\,\mu \text{m} \cdot 2.7\,\mu \text{m} \cdot \frac{3}{2} + 5.5\,\mu \text{m} \cdot 5.5\,\mu \text{m}  &= 53 \,\mu \text{m}^2 \label{SQUID_ff}
\end{align}

In order to establish a connection between the in-plane and out-of-plane rotation, we can calculate the out-of-plane component in the in-plane rotation. The magnetic field in Figure \ref{Area_SQUID} (a) oscillates with a periodicity of $\Delta B\approx 2.5mT$. Therefore according to $arccos(\frac{43\mu T}{2.5mT})=89.014^{\circ}$, the out of plane component in Figure \ref{Area_SQUID} (a) is $\approx 0.986^{\circ}$.

\section{RSJ model for a three-terminal Josephson junction}
\label{app.c}

\subsection{RSJ equations}

\begin{figure}
\centering
\includegraphics*[width=0.6\linewidth]{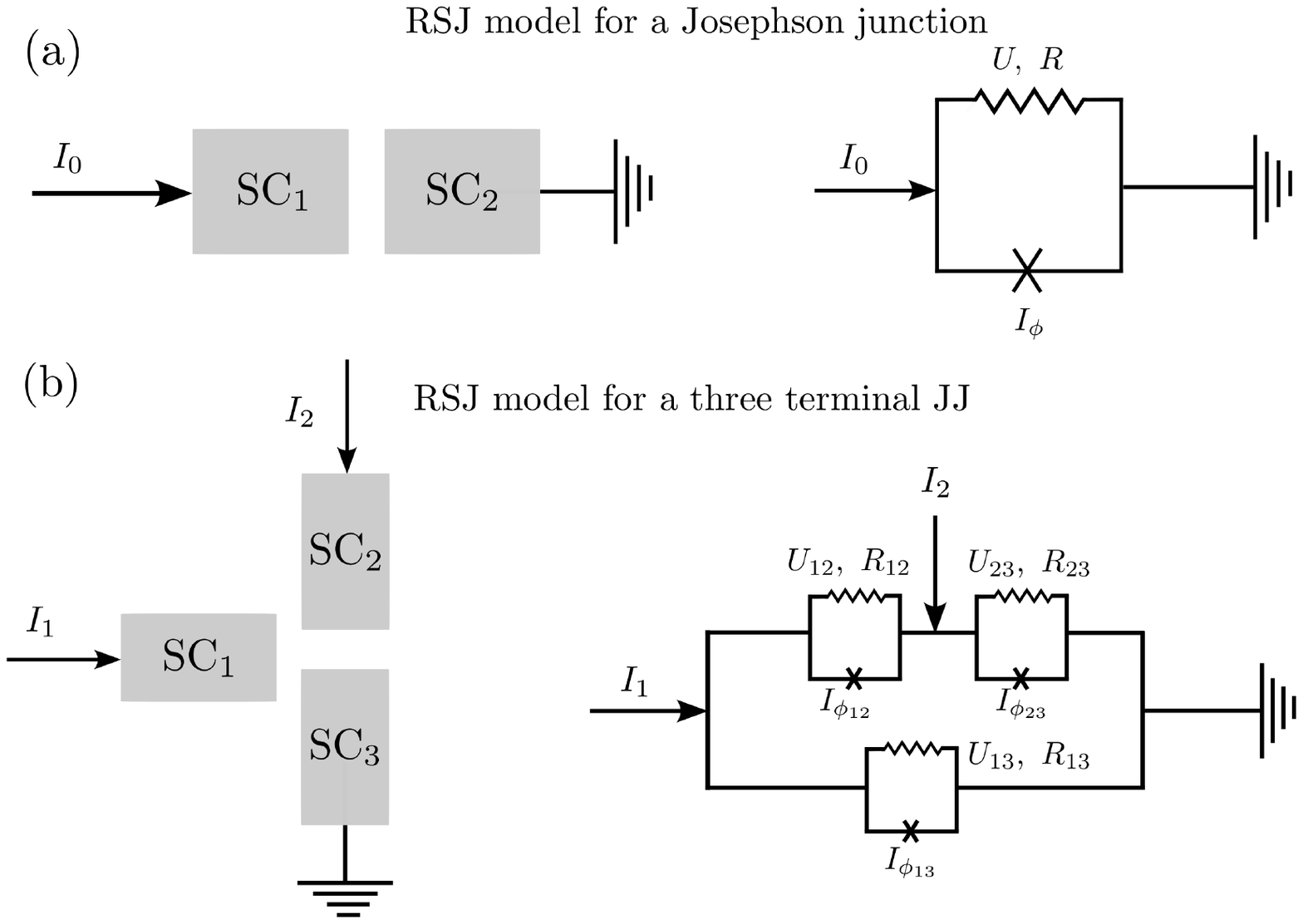} 
\caption{(a): Schematic picture of a Josephson junction and its corresponding RSJ parallel circuit, formed by a superconducting arm and a resistive arm. The junction is biased by the DC-current $I_0$.
(b): Schematic representation of a three-terminal Josephson junction, with a T-geometry. 
In this case, the system biased by means of the DC-currents $I_1$ and $I_2$ applied to the corresponding superconducting leads. Moreover, the junction is composed by two arms containing two and one RSJ circuits, with the corresponding resistances $R_{ij}$, critical currents $I_\text{c,ij}$ and the voltage generated $U_{ij}$.}\label{fig:RSJsketck}
\end{figure}

The resistively shunted junction (RSJ) model is a simplified theoretical framework used to describe the dynamics of a Josephson junction. The RSJ model is widely used to analyze the voltage-current characteristics of Josephson junctions and their response to external currents and electromagnetic fields. In its most simplified version, the Josepshon junction is represented as a parallel circuit with an ideal Josephson element in one of the arms and a normal (Ohmic) resistor in the other one, see Fig.~\ref{fig:RSJsketck}(a). Here, the resistor accounts for dissipation due to quasiparticle tunneling, while the Josephson element governs the supercurrent behavior according to the Josephson equations.  

When a bias current is applied to the JJ, both arms of the circuit are biased splits onto both arms and  
current conservation allows to relate the incoming current $I_0$ and the current flowing through the arms of the circuit, namely,
\begin{align}
I_0=I_\text{c} \sin(\phi)+\frac{U}{R}\label{eq.RSJ0_C1}
\end{align}
with the supercurrent $I_\text{c} \sin(\phi)$, the superconducting phase difference $\phi$ and the resistive current $U/R$. We can use the Josephson equation,
\begin{align}
    \frac{U}{R}=\frac{\hbar}{2eR} \frac{d}{dt}\phi.\label{eq.RSJ0}
\end{align}
to write a closed differential equation with which to calculate the phase dynamics. 

The solution of this differential equation is found, firstly, by rearranging Eq.~\eqref{eq.RSJ0}, 
\begin{align}
\frac{d}{dt}\phi=\frac{2eR}{\hbar}\left[I_0-I_\text{c} \sin(\phi)\right]   \Rightarrow \frac{d \phi}{I_0-I_\text{c} \sin(\phi)}=\frac{2eR}{\hbar}dt.
\end{align}
Then, integrating over a period $T$ where $\phi$ advances $2\pi$, that is,
\begin{align}
\int_0^{2\pi} d\phi \frac{1}{I_0-I_\text{c} \sin(\phi)}=\int_0^{T} dt \frac{2eR}{\hbar}  \Rightarrow 
\frac{2\pi}{\sqrt{I_0^2-I_\text{c}^2} }=\frac{2eR}{\hbar} T
\end{align}
Finally, using the Josephson equation that  relates the voltage and the Josephson frequency ($\omega_0$) $2eU/\hbar=\omega_0 \equiv2\pi/T$, we arrive to
\begin{align}\label{eq.V1limitI1}
 U=  R \sqrt{I_0^2-I_\text{c}^2}. 
\end{align}
The voltage-current relation following this curve shows a zero value for $|I_0|< I_\text{c}$ and a square-root relation for $|I_0|> I_\text{c}$. 

This relation has been applied to characterize a large number of experiments with Josephson junctions. Our intention now is to extend this equation to the case of a multiterminal Josephson junction and analyze the resulting regimes as a function of the bias current and magnetic fields.
To this aim, we use again current conservation on the circuit scheme of Fig.~\ref{fig:RSJsketck}~(b), and relate the external bias currents $I_1$ and $I_2$ entering through the superconducting electrodes 1 and 2 and the supercurrents $I_{\phi_{ij}}$ and resistive currents $\frac{U_{ij}}{R_{ij}}$, namely
\begin{align}
&I_1=I_\text{c,13} \sin(\phi_{13})+I_\text{c,12} \sin(\phi_{12})+\frac{U_{13}}{R_{13}}+\frac{U_{12}}{R_{12}},\label{eq.RSJ1}\\
&I_1+I_2=I_\text{c,13} \sin(\phi_{13})+I_\text{c,23} \sin(\phi_{23})+\frac{U_{13}}{R_{13}}+\frac{U_{23}}{R_{23}}.\label{eq.RSJ2}
\end{align}
with $I_{\text{c},ij}$ and $\phi_{ij}$ are the bare critical current and the phase difference between the superconducting leads $i$ and $j$.

Again, we can use the Josephson equation to rewrite the voltage generated between two superconductors and the derivative of the phase difference, that is,
\begin{align}
    \frac{U_{ij}}{R_{ij}}=\frac{\hbar}{2eR_{ij}} \frac{d}{dt}\phi_{ij},
\end{align}

\begin{figure}
\centering
\includegraphics*[width=0.65\linewidth]{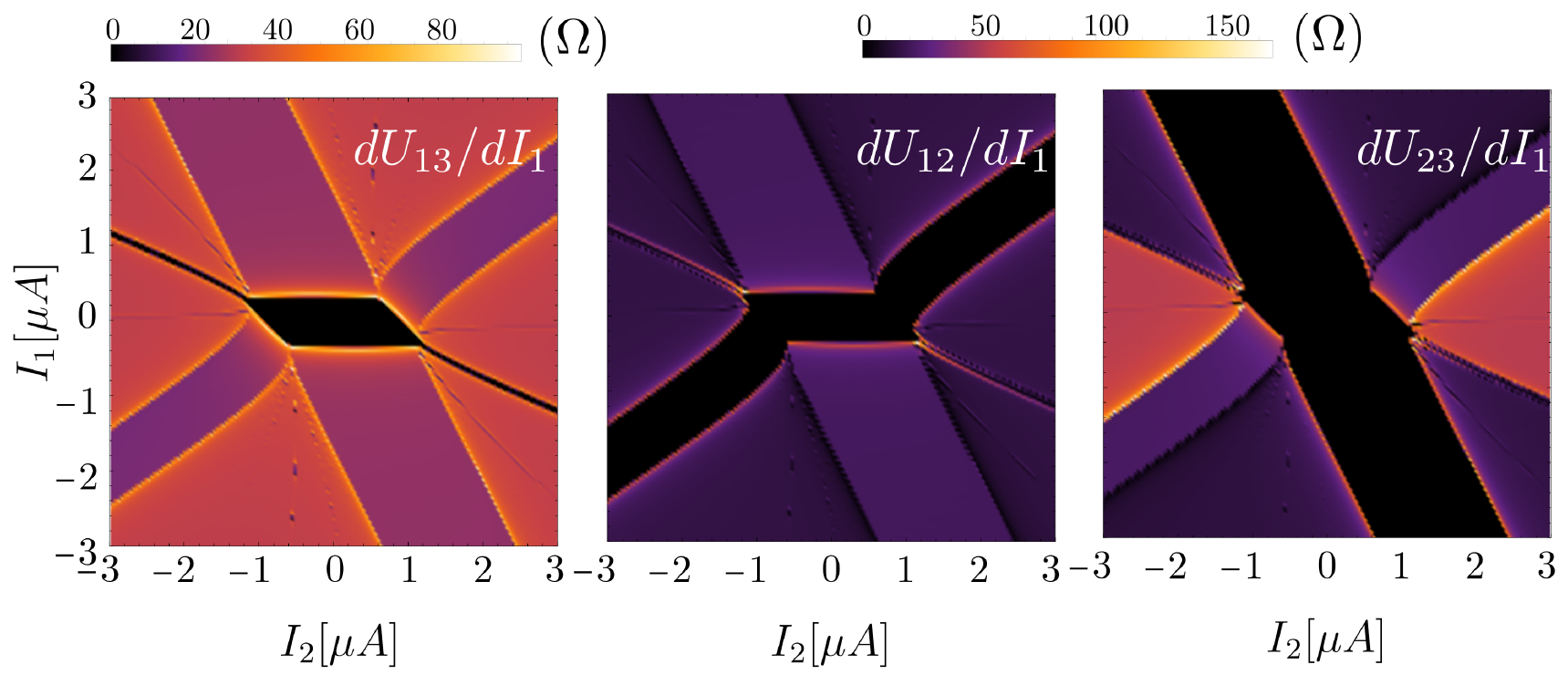} 
\caption{
Differential resistances $dU_{13}/dI_1$, $dU_{12}/dI_1$ and $dU_{23}/dI_1$ obtained from solving Eqs.~\ref{eq.RSJ1} and~\ref{eq.RSJ2}, with $I_\text{c,13}=0.055\,\mu$A, $I_\text{c,12}=0.3\,\mu$A and $I_\text{c,23}=0.9\,\mu$A and $R_{13}=53.7\,\Omega$, $R_{12}=50.7\,\Omega$ and $R_{23}=35.9\,\Omega$.}\label{fig:RSJ_I1_I2}
\end{figure}

To find a closed set of differential equations, we have to realize that the three phase differences are not independent, but related by means of 
\begin{align}
\phi_{13}=\phi_{12}+\phi_{23}.\label{eq.relphase}
\end{align}
Thus, plugging Eq.~\eqref{eq.relphase} into Eq.~\eqref{eq.RSJ2}, we eliminate $\phi_{23}$ in favor of $\phi_{13}$ and $\phi_{12}$, yielding the closed set of differential equations
\begin{align}
&I_1=I_\text{c,13} \sin(\phi_{13})+I_\text{c,12} \sin(\phi_{12})+\frac{\hbar}{2e}\frac{d}{dt}\left(\frac{1}{R_{13}}\phi_{13}+\frac{1}{R_{12}}\phi_{12} \right),\\
&I_1+I_2=I_\text{c,13} \sin(\phi_{13})+I_\text{c,23} \sin(\phi_{23})+\frac{\hbar}{2e}\frac{d}{dt}\left[\left(\frac{1}{R_{13}}+\frac{1}{R_{23}}\right)\phi_{13}-\frac{1}{R_{23}}\phi_{12} \right].
\end{align}
We solve numerically these equations, and show the differential resistances $dU_{ij}/dI_1$ as a function of the bias currents $I_1$ and $I_2$, finding a striking similarities to the experimental results, see Fig.~\ref{fig:RSJ_I1_I2}.

In contrast to the RSJ model for a single JJ, where the superconducting regime is simply set by $|I_0|<I_\text{c}$, here, the interplay of the three supercurrents and resistances makes it more difficult to predict the extension of the superconducting regimes. For this reason, in the following subsections, we analyze two bias current regimes based on how many individual JJs remain in the superconducting regime,
and try to estimate their extension as a function of the bias currents $I_1$ and $I_2$.

\subsubsection{Intermediate bias regime}

In the intermediate bias regime, two of the junctions are in the resistive regime. That is, their corresponding phases $\phi_{ij}(t)$ evolve rapidly in time, yielding an average contribution of $\langle \sin(\phi_{ij}(t))\rangle \approx 0$. We can have an analytical insight on this regime, by considering three limiting cases where only one of the supercurrents is finite, that is, we set $I_\text{c,12}=I_\text{c,23}=0$, $I_\text{c,13}=I_\text{c,23}=0$ and $I_\text{c,13}=I_\text{c,12}=0$ and obtain the average voltage generated in the junction that remains superconducting.

Let us start from the case $I_\text{c,12}=I_\text{c,23}=0$. Here, the RSJ equations given in Eqs.~\eqref{eq.RSJ1} and~\eqref{eq.RSJ2} simplify to
\begin{align}
   &I_1=I_\text{c,13} \sin(\phi_{13})+\frac{U_{13}}{R_{13}}+\frac{U_{12}}{R_{12}},\\
   &I_1+I_2=I_\text{c,13} \sin(\phi_{13})+\frac{U_{13}}{R_{13}}+\frac{U_{23}}{R_{23}}.
\end{align}
Now, this set of differential equations can be solved analytically, making again the substitution $
\phi_{23}=\phi_{13}-\phi_{12}$, and rearranging we arrive to  
\begin{align}\frac{d}{dt}
\begin{pmatrix}
\phi_{13}\\
\phi_{12}
\end{pmatrix} &= \frac{2e}{\hbar}M \begin{pmatrix}
I_1-I_\text{c,13}\sin(\phi_{13})\\
I_1+I_2-I_\text{c,13}\sin(\phi_{13})
\end{pmatrix},\label{eq.matrixform}
\end{align}
with the resistance matrix
\begin{equation}\label{eq.Mmatrix}
M = \frac{1}{R_{13}+R_{12}+R_{23}}\begin{pmatrix}
R_{13} R_{12} &  R_{13} R_{23} \\
R_{12}(R_{13}+R_{23}) & -R_{12} R_{23}\\
\end{pmatrix}.
\end{equation}
We note that the first row in Eq.~\eqref{eq.matrixform} contains only the variable $\phi_{13}$, so it allows for a separable solution. Following the same steps as in the solution for a single JJ, we arrive to
\begin{align}\label{eq.V13}
 V_{13}= \frac{R_{13}(R_{12}+R_{23})}{R_{13}+R_{12}+R_{23}} \sqrt{\left(I_1+\frac{R_{23}}{R_{12}+R_{23}}I_2\right)^2-I_\text{c,13}^2} 
\end{align}
which becomes resisitive for $\left|I_1+\frac{R_{23}}{R_{12}+R_{23}}I_2\right|>I_\text{c,13}$ and zero otherwise. As expected, this expression resembles the solution for an individual Josepshon junction, i.e.~$V=R \sqrt{I_0^2-I_\text{c}^2}$, with a modified resistance $R\rightarrow R_{13}(R_{12}+R_{23})/(R_{13}+R_{12}+R_{23})$
and bias current $I_0\rightarrow I_1+R_{23}/(R_{12}+R_{23})I_2$. 
The superconducting regime, i.e.~$V_{13}=0$ is given by a strip in the $(I_1,I_2)$-plane, with horizontal and vertical widths given by 
\begin{align}
    &\Delta I_1= 2I_\text{c,13}, \\
    &\Delta I_2= 2\frac{R_{12}+R_{23}}{R_{23}}I_\text{c,13},\\
    &\text{slope in $I_2$:~~}-\frac{R_{23}}{R_{12}+R_{23}}
\end{align}

We obtain similar results for the other two cases.  For $I_\text{c,13}=I_\text{c,23}=0$, we have
\begin{align}\label{eq.V12}
 V_{12}=  \frac{R_{12}(R_{13}+R_{23})}{R_{13}+R_{12}+R_{23}} \sqrt{\left[\frac{R_{13}}{R_{13}+R_{23}}I_1-\frac{R_{23}}{R_{13}+R_{23}}I_2\right]^2-I_\text{c,12}^2}
\end{align}
with horizontal and vertical widths given by 
\begin{align}
    &\Delta I_1= 2\frac{R_{13}+R_{23}}{R_{13}}I_\text{c,12} \\
    &\Delta I_2= 2\frac{R_{13}+R_{23}}{R_{23}}I_\text{c,12}\\
     &\text{slope in $I_2$:~~}\frac{R_{23}}{R_{13}}
\end{align}

Finally for $I_\text{c,13}=I_\text{c,12}=0$, we have
\begin{align}\label{eq.V23}
 V_{23}=  \frac{R_{23}(R_{12}+R_{13})}{R_{13}+R_{12}+R_{23}} \sqrt{\left[\frac{R_{13}}{R_{13}+R_{23}}I_1-I_2\right]^2-I_\text{c,23}^2} 
\end{align}
with horizontal and vertical width given by 
\begin{align}
    &\Delta I_1= 2I_\text{c,23} \\
    &\Delta I_2= 2\frac{R_{12}+R_{13}}{R_{13}}I_\text{c,23}\\
    &\text{slope in $I_2$:~~}-\frac{R_{12}+R_{13}}{R_{13}}
\end{align}

\begin{figure}[]
    \centering
     \includegraphics[width=0.75\textwidth, keepaspectratio] {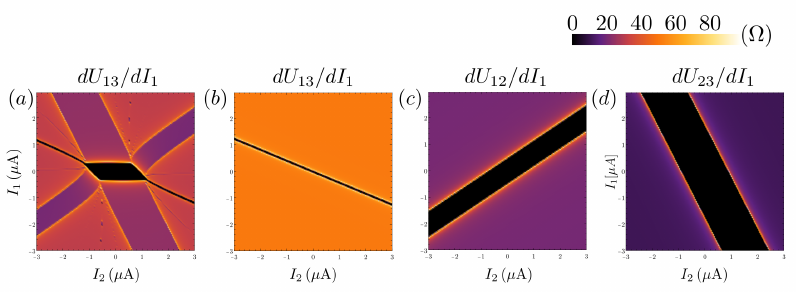}
    \caption{
   (a): Differential resistance $dR_{13}/dI_1$ as a function of the bias current $I_1$ and $I_2$, using the same parameters as in Fig.~\ref{fig:RSJ_I1_I2}. (b)-(d): Differential resistance obtained out of Eqs.~\eqref{eq.V13}, \eqref{eq.V12} and~\eqref{eq.V23}. 
    }
    \label{fig.decomposition} 
\end{figure}

The differential resistances $dU_{13}/dI_1$, $dU_{12}/dI_1$ and $dU_{23}/dI_1$ resulting from all these three limits are given in Fig.~\ref{fig.decomposition} together with the original case, where all supercurrents are finite. We can observe that the slope and extension of the 6-pointed star-like arms have a quantitatively good correspondence with the analytical calculations. 
All these results are summarized in Table~\ref{tab:interslopes} and used to estimate the value of the experimental critical current of individual Josephson junctions in the main text.

\subsubsection{Low bias regime}

\begin{figure}[]
    \centering
    \includegraphics[width=0.75\textwidth, keepaspectratio] {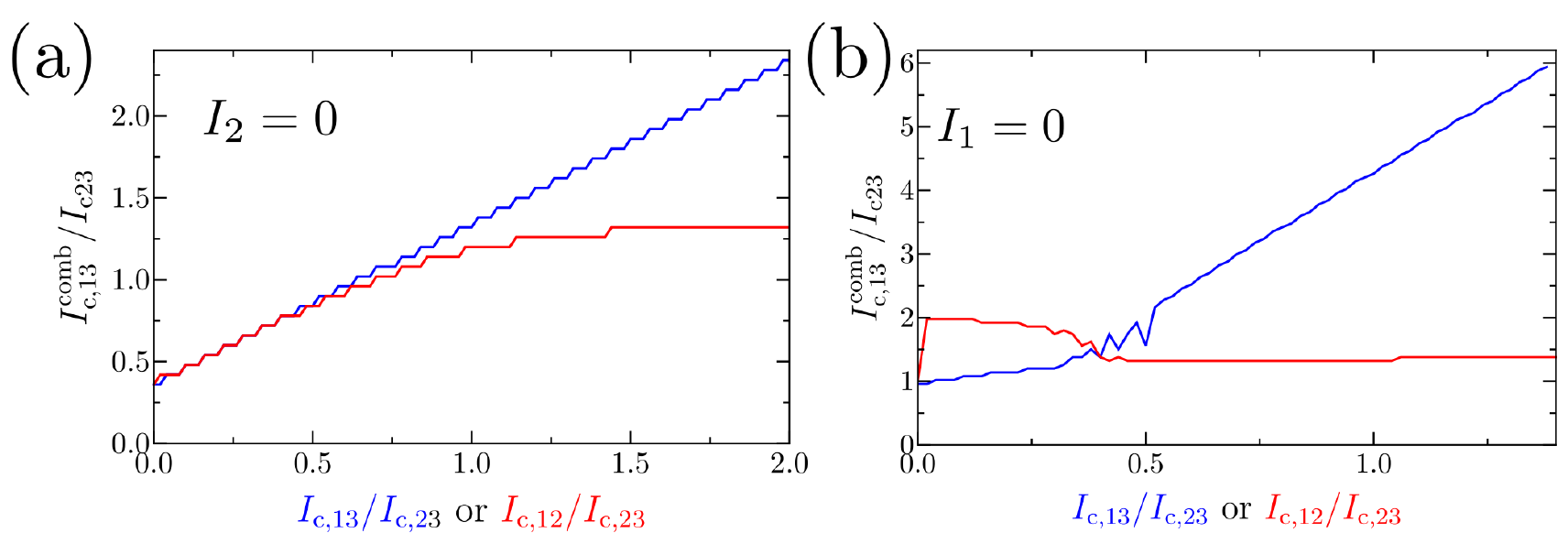}
    \caption{
    (a) and (b): Combined critical current $I_\text{c,13}^\text{comb}$ as a function of $I_\text{c,13}$ (blue curve) and $I_\text{c,12}$ (red curve) for $I_2=0$ (a) and $I_1=0$ (b).}
    \label{fig1.2} 
\end{figure}

The extension of the CCC depends on the critical current of all individual Josephson junctions and in some cases on the relative values of the resistances. In the main text, we have estimated the value of $I_\text{c,13}^\text{comb}(I_1,I_2=0)$ and $I_\text{c,13}^\text{comb}(I_1=0,I_2)$ based on an equal voltage generation in parallel circuits. Here, we provide a calculation of both combined critical currents in Fig.~\ref{fig1.2} as a function of $I_\text{c,13}$ (blue) and $I_\text{c,12}$ (red) for $I_2=0$ (a) and $I_1=0$ (b). 

For $I_2=0$, we can see that the curve behaves approximately as $I^{\text{comb}}_\text{c,13}(I_1,I_2=0)\sim I_\text{c,13}+ \text{Min}\{I_\text{c,12},I_\text{c,23}\}$. That is, as a function of $I_\text{c,13}$ the curve is linear, with slope $I^{\text{comb}}_\text{c,13}/I_\text{c,13}=1$. In turn, if we vary $I_\text{c,12}$, the curve changes linearly up to the point where $I_\text{c,12}\gtrsim I_\text{c,23}$. At this point, the curve bends and reaches a constant value of approximately $I^{\text{comb}}_\text{c,13}(I_1,I_2=0)\sim I_\text{c,13}+ I_\text{c,23}$.

For $I_1=0$ the combined critical current  $I^{\text{comb}}_\text{c,13}(I_1=0,I_2)$ exhibits a more complex behavior as a function of the individual critical currents. In particular, we can see a linear behavior with slope 1 as a function of $I_\text{c,13}$ as long as $I_\text{c,13}<I_\text{c,12}$. After that point, $I^{\text{comb}}_\text{c,13}(I_1=0,I_2)$, keeps increasing linearly but with the slope $1+R_{12}/R_{23}$. 
In general, $I^{\text{comb}}_\text{c,13}(I_1=0,I_2)$ exhibits a complicated behavior that can be summarized as, 
\[
I^{\text{comb}}_\text{c,13}(I_1=0,I_2) \approx \begin{cases}
    \left(1+\frac{R_{12}}{R_{23}}\right) I_\text{c,13} & \text{if } I_\text{c,23}< I_\text{c,13} \\
    I_\text{c,23}+\text{min}\{I_\text{c,13},I_\text{c,12}\} & \text{if } I_\text{c,23}> I_\text{c,13} 
\end{cases}
\]

\subsection{RSJ equations in the presence of a magnetic field}

\begin{figure}
\centering
\includegraphics*[width=0.6\linewidth]{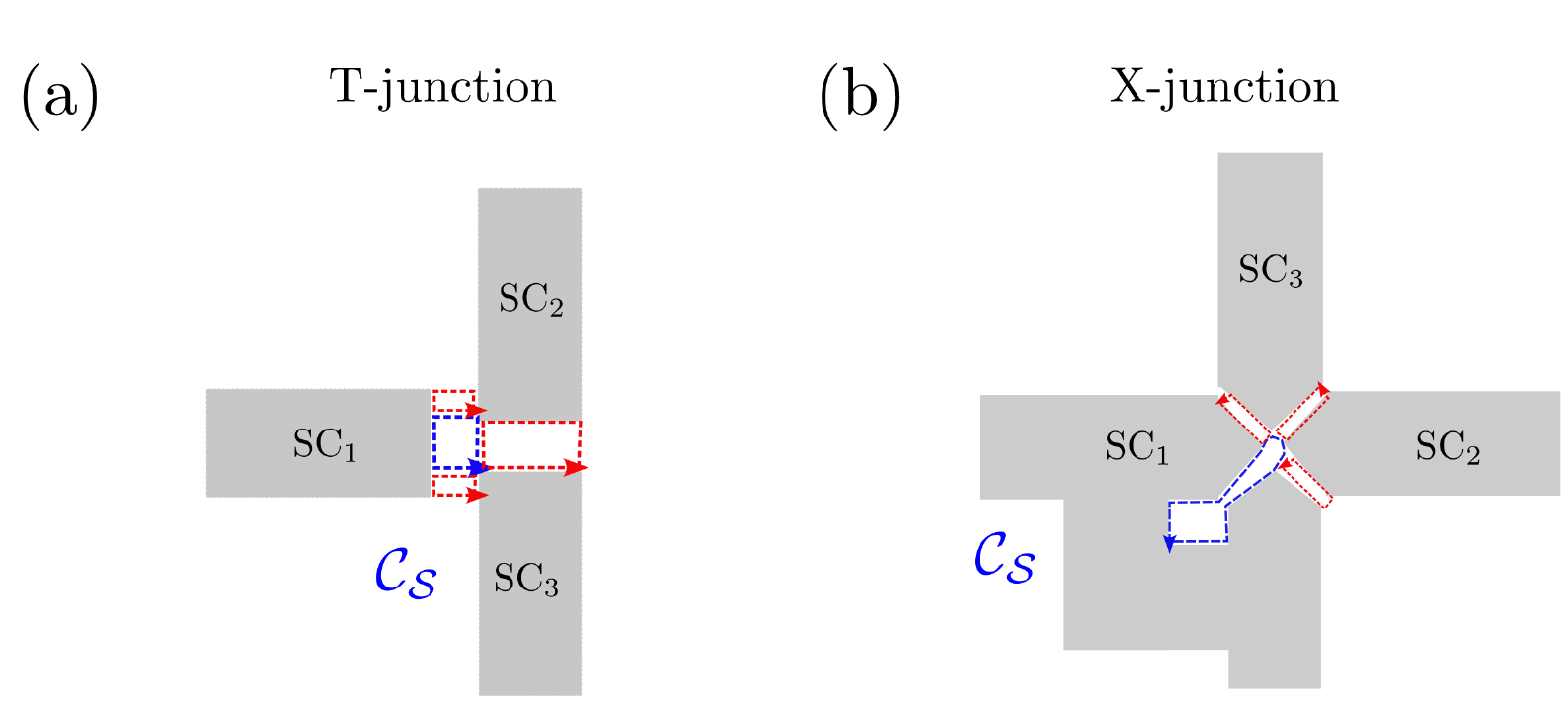} 
\caption{Schematic picture of the (a) T-shaped (b) X-shaped junction geometries with the corresponding path (blue dashed line) $\mathcal{C}_\mathcal{S}$ used in the line integrals connecting the three superconducting electrodes.}\label{fig:JJmagneticfield}
\end{figure}

We introduce the effects of a perpendicular and in-plane magnetic field in the RSJ equations derived above. To this aim, we distinguish between two types of magnetic fluxes: one that threads the normal part of each individual Josephson junction through which the supercurrent is flowing, see red paths in Fig.~\ref{fig:JJmagneticfield}. This contribution gives rise to the well-known Fraunhofer pattern. 
The second one is specific for multiterminal Josephson junctions and relates the superconducting phase differences of all junctions to the flux threading the remaining area, which was not accounted before and that links all junctions, see blue paths in Fig.~\ref{fig:JJmagneticfield}.

We begin by introducing the presence of a magnetic flux on individual Josephson junctions $\Phi_{ij}=S_{ij} B_\perp$ threading the normal part of the junction with surface $S_{ij}$ limited by the superconducting contacts $i$ and $j$, and delimited by the path $\mathcal{C}_{ij}$ (marked with a red dashed arrow in Fig.~\ref{fig:JJmagneticfield}). Assuming a constant current density profile, the supercurrent flowing between the superconducting contacts $i$ and $j$, is given by the well-known Fraunhofer pattern,  
\begin{align}
I_{\phi_{ij}}= I_{\text{c},ij} \sin(\phi_{ij}) \frac{\sin(\pi \Phi_{ij}/\Phi_0)}{\pi \Phi_{ij}/\Phi_0},
\label{eq.sinc}
\end{align}
with $\Phi_0=h/2e$ the flux quanta. 

Experimentally, the magnetic field is applied in such a way that, if finite, only a small component is out of plane, i.e.~$\vec{B}=B_\parallel \vec{e}_\parallel+B_\perp \vec{e}_\perp$, with $B_\parallel \gg B_\perp$. In this scenario, the addition of a magnetic field can suppress the pairing amplitude by providing a finite kinetic energy to the condensate, becoming energetically more favorable to reduce $\Delta$\cite{tinkham2004}, or via the superconducting Doppler effect, which reduces the pairing amplitude by tilting the quasiparticle spectrum~\cite{Tkachov2004a,Tkachov2005a, Rohlfing2009a}. The result is similar as the first lobe of the Fraunhofer pattern but without an oscillatory behavior. In this way, we replace the $\text{sinc}$ functional form on Eq.~\eqref{eq.sinc} by an exponential or gaussian suppression, namely
\begin{align}
&I_{\phi_{ij}}= I_{\text{c},ij} \sin(\phi_{ij}) F_{ij}(B_\parallel),~\text{with}  \label{eq.gauss} \\
&F_{ij}(B_\parallel)=\exp( -f_{ij} B_{\parallel}^2),
\end{align}
where $f_{ij}$ is a fitting parameter in units of mT$^{-2}$.

We now introduce the second type of magnetic flux contribution, that is, the one that is specific for multiterminal JJs. To this aim, we rewrite the RSJ equations given in Eq.~\eqref{eq.RSJ1} and Eq.~\eqref{eq.RSJ2}
\begin{align}
   &I_1=I_\text{c,13} \sin(\phi_{13})F_{13}(B_\parallel)+I_\text{c,12} \sin(\phi_{12})F_{12}(B_\parallel)+\frac{U_{13}}{R_{13}}+\frac{U_{12}}{R_{12}}\\
   &I_1+I_2=I_\text{c,13} \sin(\phi_{13})F_{13}(B_\parallel)+I_\text{c,23} \sin(\phi_{23})F_{23}(B_\parallel)+\frac{U_{13}}{R_{13}}+\frac{U_{23}}{R_{23}},
\end{align}
where we have assumed that the third electrode is grounded.

\begin{figure}
\centering
\includegraphics*[width=0.75\linewidth]{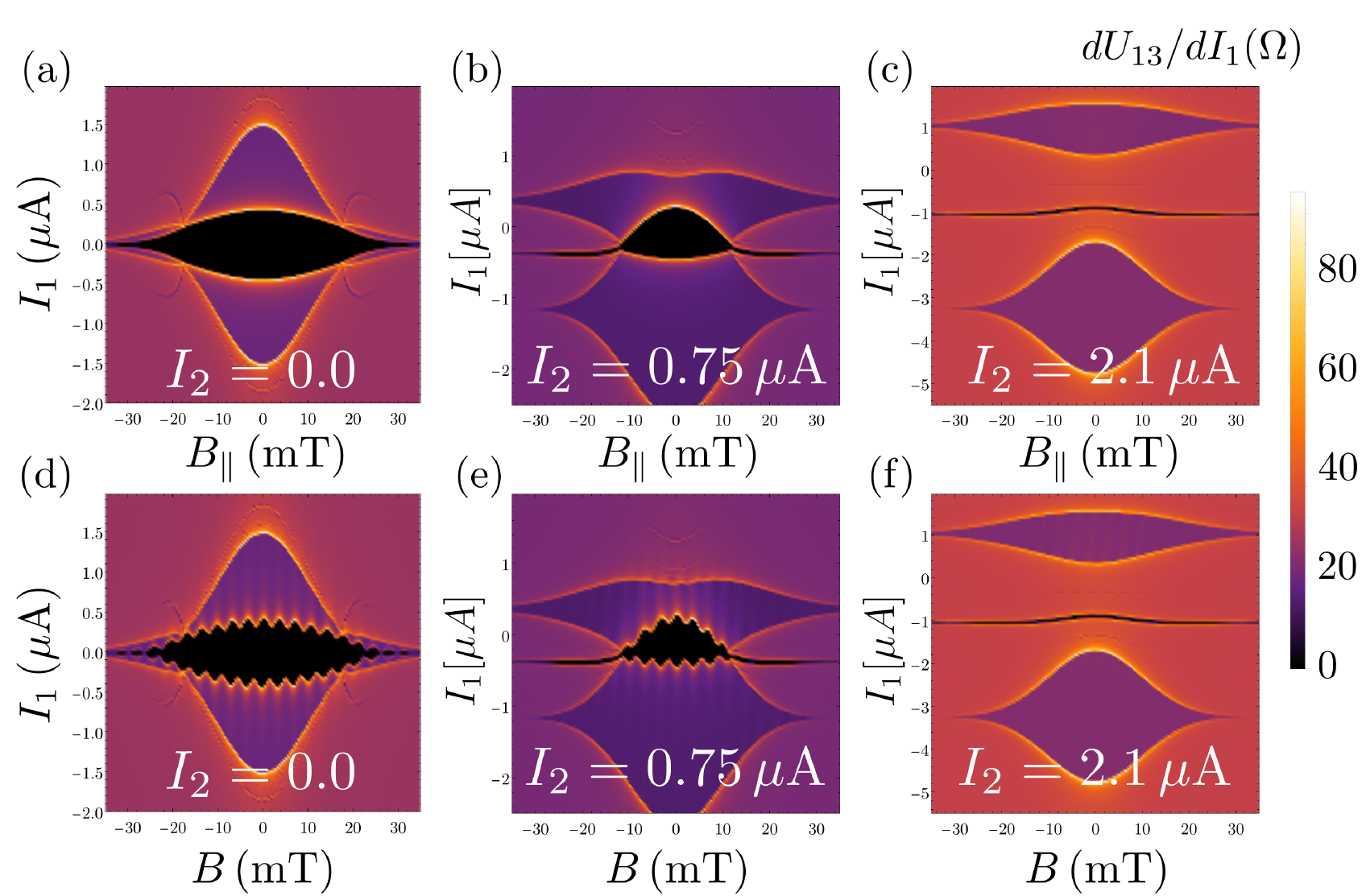} 
\caption{Differential resistance as a function of the in-plane~(a)-(c) magnetic field and a magnetic field with both in-plane and out of plane components (d)-(f). The parameters used are the same as in Fig.~\ref{fig:RSJ_I1_I2}}\label{fig:RSJ_I1_B}
\end{figure}

We now solve this set of differential equations, again, taking into account that $\phi_{13},~\phi_{12}$ and $\phi_{23}$ are not independent of each other. This time however, the magnetic flux enters into this relation via the line integral of the vector potential $\vec{A}$ generated by the magnetic field $B_\perp$ over the path $\mathcal{C}_\mathcal{S}$, which links the three junctions and encloses an area limited by $JJ_{12}$, $JJ_{23}$ and $JJ_{13}$, see Fig.~\ref{fig:RSJ_I1_I2}. Under these conditions we can write 
\begin{align}
\pi\frac{\Phi}{\Phi_0}=-\phi_{13}+\phi_{12}+\phi_{23}\Rightarrow \phi_{23}=\phi_{13}-\phi_{12}+\pi\frac{\Phi}{\Phi_0},
\end{align}
yielding the closed set of differential equations 
\begin{align}
    I_1=I_\text{c,13} \sin(\phi_{13})&F_{13}(B_\parallel)+I_\text{c,12} \sin(\phi_{12})F_{12}(B_\parallel)+\frac{\hbar}{2eR_{13}} \frac{d}{dt}\phi_{13}+\frac{\hbar}{2eR_{12}} \frac{d}{dt}\phi_{12}\label{eq:RSJ1B} \\
   I_1+I_2=I_\text{c,13} &\sin(\phi_{13})F_{13}(B_\parallel)+I_\text{c,23} \sin(\phi_{13}-\phi_{12}+\pi\Phi/\Phi_0)F_{23}(B_\parallel)\nonumber \\ 
   &+\frac{\hbar}{2eR_{13}} \frac{d}{dt}\phi_{13}+\frac{\hbar}{2eR_{23}} \frac{d}{dt}(\phi_{13}-\phi_{12}).\label{eq:RSJ2B}
\end{align}
For the T-junction we use values close to the measured experimental parameters $I_\text{c,13}=0.055\,\mu$A, $I_\text{c,12}=0.3\,\mu$A and $I_\text{c,23}=0.9\,\mu$A and $R_{13}=53.7\,\Omega$, $R_{12}=50.7\,\Omega$ and $R_{23}=35.9\,\Omega$ and the parallel magnetic field fitting parameters $f_{13}=10^{-3}$, $f_{12}=2\times10^{-3}$ and $f_{23}=5\times 10^{-3}$ in units of mT$^{-2}$. 

In Fig.~\ref{fig:RSJ_I1_B}, we show the behavior of the differential resistance $dU_{ij}/dI_1$ as a function of $I_1$ and an in-plane magnetic field $B_\parallel$~(a)-(c) and a magnetic field with both in-plane and out-of-plane components $\vec{B}=B_\parallel \vec{e}_\parallel+B_\perp \vec{e}_\perp$~(d)-(f) for different values of $I_2$. As in the experimental results, the CCC (black) restricts to the area corresponding to the area JJ$_{13}$ and the overlap between JJ$_{12}$ and JJ$_{12}$. The bias current $I_2$ shifts relatively the position of the critical current envelopes, becoming almost independent for $I_2> I_\text{c,13}^\text{comb}(I_1=0, I_2)$, introduced above. As we anticipated, the presence of $B_\parallel$ reduces the critical current contribution of all junctions. If an out-of-plane magnetic field component is present, a SQUID-pattern behavior arises only at the CCC areas, where all JJs coincide, see Fig. \ref{fig:RSJ_I1_B}~(d)-(e). For a large bias current $I_2$, all JJs become effectively decoupled, and consequently, the SQUID-pattern is lost, resulting in the same behavior as just having an in-plane magnetic field, see Fig. \ref{fig:RSJ_I1_B}~(c) and~(f).

\end{appendix}

\end{document}